\documentclass[%
 reprint,
superscriptaddress,
 amsmath,amssymb,
 aps,
]{revtex4-2}

\usepackage{graphicx}
\usepackage{dcolumn}
\usepackage{bm}
\usepackage{appendix}
\usepackage{hyperref}
\usepackage{color}

\hypersetup{
  colorlinks   = true, 
  urlcolor     = black, 
  linkcolor    = blue, 
  citecolor   = blue 
}

\begin{document}

\title{Conductance of correlated many-fermion systems from charge fluctuations}

\author{Yuchi He}
\affiliation{Institut f{\"u}r Theorie der Statistischen Physik, RWTH Aachen University and
  JARA---Fundamentals of Future Information Technology, 52056 Aachen, Germany}
\author{Dante M.~Kennes}
\affiliation{Institut f{\"u}r Theorie der Statistischen Physik, RWTH Aachen University and
  JARA---Fundamentals of Future Information Technology, 52056 Aachen, Germany}
 \affiliation{Max Planck Institute for the Structure and Dynamics of Matter, Center for Free Electron Laser Science, 22761 Hamburg, Germany}
\author{Volker Meden}
\affiliation{Institut f{\"u}r Theorie der Statistischen Physik, RWTH Aachen University and
  JARA---Fundamentals of Future Information Technology, 52056 Aachen, Germany}

\date{\today}

\begin{abstract}
We put forward a relation between the static charge fluctuations and the conductance of correlated many-fermion systems at zero temperature, avoiding the use of time-dependent fluctuations as in the fluctuation-dissipation theorem. Static charge fluctuations can efficiently be computed for low-dimensional systems using tensor network approaches, while the conductance is often significantly more difficult to  obtain, requiring a challenging low-frequency linear response computation or an explicit time evolution. We put this relation to the test for quantum dot and quantum point contact setups, where in limiting cases exact results are known. Our study includes systems in which the one-dimensional reservoirs are interacting.    
\end{abstract}

\maketitle


\section{Introduction}\label{intro}
The extreme miniaturization of electronic devices is fascinating from both, the academic as well as the engineering perspective. In research laboratories, the limit of low-temperature, single-channel quantum transport can nowadays routinely be reached and the linear electronic transport in this limit can be characterized by a gate-controllable conductance of the order of the conductance quantum. Numerous theoretical studies were motivated by this progress. One hallmark is the Landauer–Büttiker formula, which relates the conductance to scattering processes~\cite{PhysRevLett.46.618, PhysRevLett.57.1761}. 

On the one hand, the computation of scattering characteristics of free fermions in the presence of external potentials (single-particle scattering)~\cite{groth2014kwant} was used extensively to compare with interesting experimental observations of the linear conductance. On the other hand, two-particle interactions between the fermions can lead to even richer phenomena. However, in the presence of such interactions, fermionic single-particle scattering theory can, in general, no longer be applied. Using ideas from scattering theory also for interacting systems thus always requires additional reasoning. For example, a scattering approach employing collective plasmons was implemented for the Tomonaga-Luttinger model~\cite{PhysRevB.52.R17040}.

Transport phenomena related to many-body Kondo physics~\cite{hewson} were observed in quantum dots coupled to two- or higher-dimensional leads~\cite{Goldhaber-Gordon1998,doi:10.1126/science.289.5487.2105}. In such setups, usually only the local on-dot two-particle interaction matters, and the electrons in the leads can be considered as noninteracting. A gate voltage controlled few-level quantum dot constitutes a comparably simple electronic device, which, however, is still of technological relevance~\cite{RevModPhys.64.849}. In another class of systems, a (simple) device is coupled to quasi-one-dimensional wires acting as the leads. In such systems, electronic correlations have a strong effect even in the leads and imply another state of matter, the Tomonaga-Luttinger liquid~\cite{Giamarchi_book}. In linear transport this, e.g., shows in the phenomena of charge fractionalization~\cite{PhysRevB.52.R17040} and universal impurity scaling~\cite{PhysRevB.9.2911,PhysRevLett.68.1220,PhysRevB.46.15233,Meden_2008}. Several experiments were interpreted in the light of these theoretical predictions~\cite{Steinberg2008, Briggeman769,doi:10.1126/science.1061797}. Related experimental setups were modeled and predictions beyond linear transport were made in Refs.~\cite{PhysRevLett.91.266402,PhysRevLett.92.226405,PhysRevB.71.165309}.
Alternatively, the physics of a single impurity in a Tomonaga-Luttinger liquid can experimentally be realized (emulated) in dissipative quantum circuits~\cite{PhysRevLett.93.126602,Mebrahtu2012,Jezouin2013,refId0}. Also, the famous 0.7 anomaly, observed in transport through a quantum point contact (QPC)~\cite{PhysRevLett.77.135,PhysRevLett.88.226805}, constituting a more complex device as compared to a quantum dot, was suggested to result from local two-particle interactions~\cite{PhysRevB.64.155319,PhysRevLett.89.196802,PhysRevB.70.245319,Bauer2013}.

To study the linear response transport of many-fermion systems in the presence of local as well as bulk two-particle interactions and to make quantitative predictions for microscopic models, various types of numerical methods for calculating the conductance were developed. Examples include the numerical renormalization group (NRG)~\cite{RevModPhys.80.395, NRGMPS}, the quantum Monte Carlo approach~\cite{QMCO, PhysRevLett.116.036801, QMC2}, and matrix product state (MPS)-based methods~\cite{Ueda,PhysRevLett.88.256403,WhiteFeiguin,schneider2006conductance,PhysRevB.79.235336,PhysRevLett.101.140601, PhysRevB.73.195304, andp2010, PhysRevLett.124.137701, kang2021twowire, EPL,PhysRevB.75.241103,PhysRevB.96.195111,bischoff2019density, PhysRevB.67.193303,PhysRevLett.105.226803,PhysRevB.85.045120, PhysRevB.86.075451,NRGMPS}. While NRG is very accurate for systems with a few interacting degrees of freedom, e.g., quantum dots, the MPS-based methods are favorable for solving models with a larger number of interacting fermions.

\begin{figure}
    \centering
    \includegraphics[width=\columnwidth]{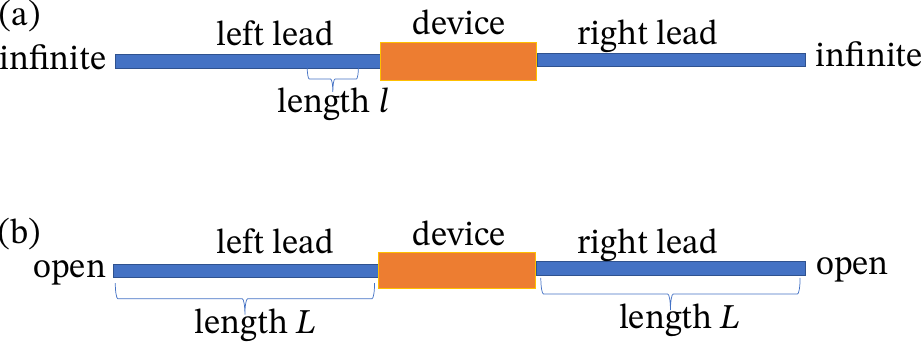}
    \caption{Sketch of the setups considered in the infinite system and the finite system geometries to compute the charge fluctuations. }
    \label{fig:cvblock}
\end{figure}

\begin{figure}
    \centering
    \includegraphics[width=\columnwidth]{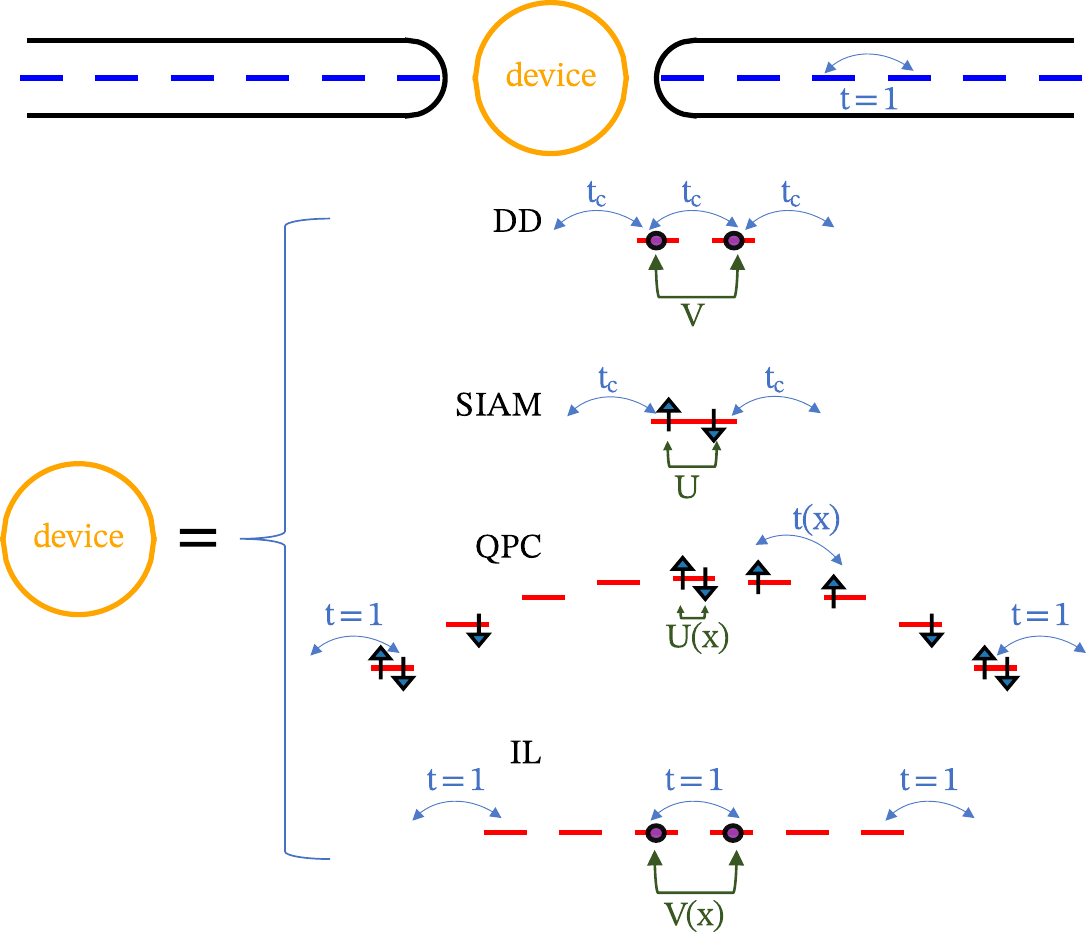}
    \caption{Sketch of the setup investigated consisting of two one-dimensional leads and a device. The fermions in the combined system are taken as either spinless or spinful, as indicated by the bullets (spinless) or arrows (spinful) in the sketches of the corresponding devices. The blue (red) short, horizontal lines in the leads (devices) denote lattice sites. The leads are taken as non-interacting for the first three devices and as interacting (nearest-neighbor interaction) in the last. The hopping matrix element $t$ in the leads is set to 1 and sets our energy scale. The device DD is a model for a spinless double quantum dot with nearest-neighbor interaction $V$ which is tunnel coupled by a hopping $t_{\rm c}<t$ to the leads. The device SIAM is the single-impurity Anderson model with tunnel coupling $t_{\rm c}$. Besides the on-site interaction $U$, the dot site is subject to a Zeeman field and a gate voltage; not indicated. The device QPC is a model for a spinful quantum point contact. The heights of the QPC lattice sites represent their on-site energies. The on-site interaction $U(x)$ and the hopping $t(x)$ of the QPC are site/bond dependent. The device IL is a model for the interface of two spinless interacting leads. In the interface region, the two-particle interaction $V(x)$ is spatially dependent (inhomogeneous).}
    \label{fig:models}
\end{figure}

In this paper, we focus on the zero-temperature linear response (zero-bias) dc conductance of devices modeled as one-dimensional chains which are coupled to two homogeneous one-dimensional leads. We are eventually interested in the case of semi-infinite leads (denoted as transport geometry) but for technical reasons have to work with large finite ones and extrapolate; see Fig.~\ref{fig:cvblock}. 
The fermions in the leads can either be noninteracting or interacting. We consider different devices (quantum dots and QPCs) which are spatially structured, i.e., they are inhomogeneous; see Fig.~\ref{fig:models} for sketches. In most examples, we take the fermions on the device part of our system to be interacting. We suggest a way to extract the linear conductance from a static ground-state property, the charge fluctuations, of a setup in which the two leads have open boundary conditions (OBCs) on the ends opposite to the device; see Fig.~\ref{fig:cvblock}(b). Time-dependent fluctuations were extensively studied, for example, in Refs.~\cite{PhysRevB.84.205129, PhysRevLett.75.2196}. Our study contributes to revealing relations between static correlations/fluctuations and transport coefficients~\cite{PhysRevLett.105.226803,PhysRevB.85.045120, kang2021twowire}, which go beyond the more conventional relations between time-dependent fluctuations and dissipation.

\textit{A priori} we expect the
static charge fluctuation approach to lead to much lower computational costs
than algorithms for dynamical properties. Computing static properties only requires the MPS-based ground-state density matrix renormalization group (DMRG)
algorithm~\cite{White}, which is highly efficient and reliable.
In comparison, for dynamical properties, computationally more costly  algorithms have to be applied, e.g., MPS-based time-evolving block decimation (TEBD) and time-dependent variational principle (TDVP)~\cite{frequencyDMRG, PhysRevB.66.045114, TEBD,TDVP}. These algorithms provide methods to numerically extract the linear conductance from the computation of a dynamical correlation function or from the finite-bias nonequilibrium steady-state current in the limit of small bias voltages.
The latter approach requires computing the current as a function of time for times larger than a transient time scale after which the current becomes steady. We refer to it as the time-dependent approach.  By explicit comparison of results obtained by the time-dependent approach and from the charge fluctuations for several examples, we here confirm that the latter can be used to obtain the zero-temperature conductance. We also exemplify the computational advantage of the charge fluctuation approach. 

While the dc conductance is defined as a dynamical quantity (see below for details), the Landauer formula, relating conductance to the transmission probability, points to a way to determine the conductance from a static property.  We propose and examine a relation between the conductance and the scaling of the static charge fluctuations in the ground state. This idea is based on previous observations and results. In a case study, a one-dimensional microscopic free fermion model with a bond impurity was examined~\cite{Peschel_2005, andp, Peschel_2012}. It was analytically shown that ground-state R\'enyi entropies diverge logarithmically as a function of the system size, with the prefactor depending on the transmission probability and thus the conductance. A similar dependence was obtained in the finite-particle number scaling of the charge fluctuations in a simplified continuum model~\cite{PhysRevLett.107.020601, Calabrese_2012A, Calabrese_2012B}. Furthermore, a study~\cite{Song} of the charge fluctuation scaling of a system of homogeneous interacting spinless fermions indicated that the prefactor of the logarithmic divergence is proportional to the Tomonaga-Luttinger liquid parameter $K$, which equals the conductance of the homogeneous system~\cite{Giamarchi_book}.   

Here, we study more general situations. In Sec.~\ref{sec1}, we first provide general arguments and analytical evidence for a relation between the finite-size scaling of the charge fluctuations and the zero-temperature conductance for different setups. We prove that this relation holds for a generic class of fermion models with noninteracting leads and a noninteracting device. We argue that the reasoning can be generalized to interacting devices coupled to noninteracting leads. We work with an effective continuum theory to heuristically reason that the relation can also be extended to interacting leads. In Sec.~\ref{numericalmethod}, we briefly summarize the well-established numerical approaches we use to compute the charge fluctuations and the current in the limit of small bias voltages. We also discuss the numerical costs of the different MPS-based methods.

We investigate several microscopic models in Sec.~\ref{sec2}. They are sketched in Fig.~\ref{fig:models}. Three setups of noninteracting leads connected to different interacting devices are considered first in Sec.~\ref{noninteractingleads}. The first two are minimal models for spinless and spinful fermions and the device is a simple quantum dot. Our spinless dot model shows the correlation physics of the so-called interacting resonant level mode, see, e.g. Ref.~\cite{PhysRevLett.101.140601} and references therein, while the spinful dot model shows the Kondo effect~\cite{hewson}. For both models, the values of the conductance extracted from charge fluctuation data are compared to those obtained from the time-dependent method. Within the error bounds of the respective approaches, agreement is found. For the spinless model, an interesting relation between the effective central charge and the conductance is examined in addition. As the third setup, an example for a more complex interacting device coupled to noninteracting leads is studied by modeling an interacting QPC and considering spin-1/2 fermions. This model is employed to understand the 0.7 anomaly, in which spin is expected to play a role~\cite{PhysRevLett.77.135,PhysRevLett.88.226805}. We show that employing the charge fluctuation method, we are able to extract a reliable estimate for the conductance while the time-dependent reference approach can no longer be used due to computational obstacles related to the complexity of the model. 

Further increasing the number of interacting degrees of freedom, in Sec.~\ref{interactingleads}, we consider devices with two leads of interacting fermions. This setup might, e.g., be relevant to model the contact resistance induced at the interface of two nanowires possibly made from different materials. Already the spinless case shows interesting correlation effects. The two-particle interaction in the two leads is allowed to differ; this leads to interesting effects and might be realized in experiments~\cite{PhysRevB.85.045120,conductingfixedpoint1,conductingfixedpoint2,kang2021twowire}. In the vicinity of the device, the two-particle interaction changes either smoothly (adiabatically) or abruptly in space, however, the bulk part of the leads is homogeneous. The device is given by a few lattice sites with variable single-particle parameters (hopping matrix elements, on-site energies). If the change of the two-particle interaction is sufficiently smooth and the device is removed, we expect from field theoretical considerations~\cite{PhysRevB.52.R8666,PhysRevB.52.R5539,PhysRevB.52.R17040,contact} that the conductance is given by an effective Tomonaga-Luttinger liquid parameter $K_{\mathrm{eff}}$ which depends on the respective parameters of the two leads. We confirm this in Sec.~\ref{interactingleads_smooth} by our suggested charge fluctuation method as well as by the conventional finite-bias time-dependent simulations. In the following, we refer to this situation as the maximal conductance fixed point. 

One direction that deserves further investigation, and which is tackled in Sec.~\ref{interactingleads_tuning}, is that of corrections due to the nonadiabaticity of the two-particle interaction when varied from the bulk value of the left lead to the one of the right lead. It was shown that, similar to single-particle backscattering~\cite{PhysRevB.9.2911,PhysRevLett.68.1220,PhysRevB.46.15233}, this can lead to a vanishing conductance~\cite{contact}. For small to intermediate two-particle interactions, signatures of this can, however, only be found on exceedingly small energy scales. Based on tree-level renormalization group (RG) arguments, it was conjectured~\cite{PhysRevB.86.075451, conductingfixedpoint1, conductingfixedpoint2} that, in the presence of single-impurity inhomogeneities and a $K_{\mathrm{eff}}>1 (<1)$, one generically obtains a maximal (zero) conductance fixed point. Furthermore, in Refs.~\cite{conductingfixedpoint1, conductingfixedpoint2} it was argued that the total backscattering is affected by the interaction through the effective Fermi velocity which is renormalized by the interaction. Given this, fine-tuning of single- and two-particle terms can lead to a maximal conductance fixed point even for $K_{\mathrm{eff}}<1$, due to an accidentally vanishing of the total backscattering.  

In Sec.~\ref{interactingleads_tuning}, we use our method to study the effect of nonadiabatically varying interactions on the conductance. With our approach, we can reach a comparably low energy scale, an order of magnitude smaller than the one obtained in Ref.~\cite{conductingfixedpoint2}.  We focus on the case $K_{\mathrm{eff}}=1$, for which backscattering is not renormalized at the tree level.  We consider the case of two interacting leads that are linked through a device with a few noninteracting sites.  We find that, on the one hand, the computed conductance is almost maximal for an odd number of device sites. On the other hand, it is significantly smaller for an even number of sites. The corresponding conductance values are found to be well estimated by the formula of the backscattering parameter for weak interactions given in Ref.~\cite{conductingfixedpoint1,conductingfixedpoint2}. Our results show that the tree-level theory works well in our microscopic setup and for the energy scales we can reach. Our method could be useful to explore the validity of the tree-level prediction for $K_{\mathrm{eff}} \neq 1$ in the future.

\section{Computing the conductance from charge fluctuation scaling}\label{sec1}

In this section, we primarily discuss general arguments and analytical insights suggesting a relation between the zero-temperature conductance in an infinite system transport geometry and the finite-size scaling of the charge fluctuations in a corresponding setup with finite leads. We propose that the charge fluctuations can be computed for a system with OBCs of the leads at the ends opposite to the device, which is particularly suitable for the highly efficient DMRG algorithm~\cite{White, RevModPhys.77.259,SCHOLLWOCK201196}. In addition to charge fluctuation scaling, we briefly discuss finite-size scaling of the entanglement entropy which can also be employed to obtain the conductance.

\subsection{The conductance and charge fluctuations}

The linear response conductance $g$ can be defined via the equilibrium current-current correlation function of a system with semi-infinite leads, as sketched in Fig.~\ref{fig:cvblock} (a). A dimensionless $G$ follows by taking the ratio of $g$ and the conductance quantum $e^2/h$. Employing the Kubo formula~\cite{PhysRevB.52.R5539}, one obtains
\begin{align}~\label{gkubo}
G \equiv \frac{g}{e^2/(2\pi\hbar)}=\lim_{\omega \rightarrow 0}\frac{2\pi}{\omega}\int_{0}^{\infty} dt e^{i\omega t}\left \langle [j(x,t) ,j(y,0)] \right \rangle,
\end{align}
where $x,y$ denotes an arbitrary pair of positions or lattice sites and $j$ is the  current operator. The lattice models we consider only have nearest-neighbor hoping and $j(x)=-i t(x)(c^{\dagger}_xc_{x+1}-c^{\dagger}_{x+1}c_x)$. Here $t(x)$ is the hopping matrix element multiplied to the nearest-neighbor hopping term $\sim c^{\dagger}_xc_{x+1}+{\rm H.c.}$ (kinetic energy density) of the Hamiltonian and $c^{(\dagger)}_x$ are the fermionic ladder operators (standard notation).  

We reiterate that we exclusively focus on the zero-temperature limit, where the expectation value in the current-current correlation function is taken in the ground state.  

All the models we consider have a homogeneous nearest-neighbor hopping in the leads. The amplitude of this we set to 1. Energies are measured in units of this bulk hopping and thus become dimensionless. Time is measured in units of the inverse bulk hopping and thus becomes dimensionless as well. 

The definition Eq.~(\ref{gkubo}) of the conductance requires the computation of an equilibrium dynamical correlation function of a system with semi-infinite leads. Alternatively, the linear conductance can be computed as
\begin{align}
G = \lim_{V_{\mathrm{bias}} \to 0} \frac{2\pi I_{\rm st}}{V_{\mathrm{bias}}}, 
\label{galt}
\end{align}
with $I_{\rm st} = \lim_{t \to \infty} \left< j(x,t)\right>$ being the nonequilibrium steady-state current obtained in the presence of a finite bias voltage (and semi-infinite leads). The asymptotic, steady-state value is independent of the position (bond) $x$. To obtain $I_{\rm st}$, one can study the long-time limit of a setup in which the bias voltage is switched on at time $t=0$ (quench dynamics).  

We aim to show that at zero temperature, $G$ can also be obtained by a finite-size scaling of a time-independent equilibrium quantity, namely, the charge fluctuations $\left\langle N_{p}^2 \right\rangle-\left\langle N_{p} \right\rangle^2$, where $N_p=\sum_{x \in p} n_x$ and $n_x$ is the particle number operator on site $x$. Here $p$ denotes a certain subsystem of the system of interest. The main merit of our paper is to provide concrete numerical evidence for the proposed relation of charge fluctuations and conductance. However, to first plausibilize the relation, we provide a reasoning which progresses along with three steps which we describe next.

(i) For an infinite metallic system, see Fig.~\ref{fig:cvblock} (a), the equal-time fluctuations of the total charge within one lead are expected to diverge. This follows from the large $|x-y|$ behavior of the connected two-point correlation function $ \left\langle n_{x}n_{y} \right\rangle- \left\langle n_{x} \right\rangle \left \langle n_{y} \right\rangle $ and
\begin{align}~\label{variancefromcorrelation}
\left\langle N_{p}^2 \right\rangle-\left\langle N_{p} \right\rangle^2=\sum_{x,y \in p} [\left\langle n_{x}n_{y} \right\rangle- \left\langle n_{x} \right \rangle \left \langle n_{y} \right\rangle],
\end{align}
where $p$ is now the entire semi-infinite left or right lead. We expect that $[\left\langle n_xn_y \right\rangle- \left\langle n_x \right \rangle \left \langle n_y \right\rangle]$ contains terms decaying as $(x-y)^{-2}$ for large $|x-y|$. Roughly speaking, summing over $x$ yields the asymptotic decay $\sim y^{-1}$  for large  $|y|$; further summing over $y$ implies the divergence. A truncation of the summation yields a term which diverges logarithmically with the cutoff. This observation leads to the following intuition. Restricting both reservoirs to finite leads of length $L$ and taking OBCs at the two ends opposite to the device, see Fig.~\ref{fig:cvblock} (b), the charge fluctuations is expected to diverge $\sim \ln(L)$. One of the central points of this paper is to provide evidence that the prefactor of the $\ln(L)$ divergence in this finite-size system (not a transport geometry anymore) encodes $G$
\begin{align}~\label{cvscaling}
\left\langle N_{p}^2 \right\rangle-\left\langle N_{p} \right\rangle^2 = \frac{G}{2\pi^2}\ln(L)+ \ldots ,
\end{align}
where the dots denote terms which for $L \to \infty$ do not diverge. 

We note in passing that if a conserved spin polarization $S_z$ exists, we expect a similar relation to hold for the spin conductance. 

(ii) To provide an analytical justification of Eq.~\eqref{cvscaling} in step (iii) below, it is convenient to consider an alternative setup to compute the charge fluctuation scaling. The subsystem $p$ is now chosen as a segment of length $l$ in a semi-infinite lead as sketched in Fig.~\ref{fig:cvblock} (a). We thus return to a transport geometry. The end of the subsystem close to the device is fixed. Let us assume that we were able to analytically show that 
\begin{align}~\label{cvscalinginf}
\left\langle N_{\mathrm{seg}}^2 \right\rangle-\left\langle N_{\mathrm{seg}} \right\rangle^2 = \frac{G_{\mathrm{h}}+G}{2\pi^2}\ln(l)+ \ldots ,
\end{align} 
where $G_{\mathrm{h}}$ is the conductance of the uniform infinite lead. This would complete our reasoning as Eq.~\eqref{cvscaling} can be justified based on Eq.~\eqref{cvscalinginf} employing charge conservation and the locality of charge fluctuations. The fluctuations within the segment of length $l$ stem from its two ends.  For sufficiently large $l$, the fluctuations at the end opposite to the device should behave very similar to those of a homogeneous infinite chain. At this end, the charge fluctuations thus probe the homogeneous case and there is evidence~\cite{Song}  that they are  given by $\frac{G_{\mathrm{h}}}{2\pi^2}\ln(l)$. 
Subtracting this contribution from Eq.~\eqref{cvscalinginf}, we obtain the contribution from the end close to the device, which is $\frac{G}{2\pi^2}\ln(l)$.

(iii) What remains to be argued in favor of is Eq.~\eqref{cvscalinginf}. For our analytical reasoning on this, we focus on single-channel, spinless systems.

(iiia) We first consider systems with noninteracting leads.  In this case, $G_{\mathrm{h}}=1$ in Eq.~\eqref{cvscalinginf}.  If, in addition, the device is noninteracting, exact results can  be obtained employing single-particle scattering theory.  In a transport geometry with two semi-infinite leads, the annihilation operator of a fermion located on lattice site $x$ can be written in terms of a scattering basis. The latter is obtained by solving the single-particle Hamiltonian of the complete system. For $x$ in the left lead, neglecting possible bound state(s), which neither contribute to the conductance nor the charge fluctuation scaling, the local annihilation operator can be written as 
\begin{align}
c(x) = \int_{0}^{\pi} \frac{dk}{\sqrt{2\pi}} & \left[  \left(e^{ikx}+r_{\mathrm{L}}(k)e^{-ikx}\right) c_{\mathrm{L}}(k) \right. \nonumber \\
& \left. + t_{\mathrm{R}}(k) e^{-ikx}c_{\mathrm{R}}(k) \right] ,   
\label{scat_state}
\end{align}
where $c_{\mathrm{L}}(k)$ [$c_{\mathrm{R}}(k)$] annihilates a left (right) scattering state with momentum $k>0$ and $t_{\mathrm{L},\mathrm{R}}(k)$ and $r_{\mathrm{L},\mathrm{R}}(k)$ are the transmission and reflection amplitude, respectively. The  Laudauer-B\"uttiker formula gives $G=|t_{\mathrm{R}}(k_\mathrm{F})|^2$. In Appendix \ref{app_A},  we compute the charge fluctuations employing scattering states and show that 
\begin{align}
\left\langle N_{\mathrm{seg}}^2 \right\rangle-\left\langle N_{\mathrm{seg}} \right\rangle^2 = \frac{1+G}{2\pi^2}\ln(l) + \ldots .
\label{charge_fluc}
\end{align}

If we ease the condition of a noninteracting device, single-particle scattering theory can no longer be applied directly. However, earlier studies of the zero-temperature conductance indicate that the conductance can be computed from an effective transmission amplitude~\cite{Oguri,PhysRevB.64.155319,PhysRevB.67.193303}. This is linked to the vanishing of current vertex corrections~\cite{Oguri}. We expect that this transmission amplitude also enters the charge fluctuations in exactly the same way as for a noninteracting device. However, the ultimate verification of the relation Eq.~(\ref{charge_fluc}) between the conductance and finite-size charge fluctuation scaling for interacting devices comes from the numerical evidence we will present in Sec.~\ref{noninteractingleads}. 

(iiib) We second consider setups with interacting leads. The low-energy physics of an isolated homogeneous interacting metallic one-dimensional wire is described by the Tomonaga-Luttinger model~\cite{Giamarchi_book}, a continuum field theory, and is characterized by a single Tomonaga-Luttinger parameter $K$. For a noninteracting lead $K=1$, for a repulsive bulk interaction $K<1$ and for an attractive one $K>1$. We study the case with possibly different $K$ for the two leads: $K_{\mathrm{L}}$ for the left and $K_{\mathrm{R}}$ for the right. We employ the bosonization approach~\cite{Giamarchi_book} and use the notation that the density is given by $n(x)=\bar{n}+\partial_x \phi(x)/\pi+ \ldots$, where $\bar{n}$ is the average density and  $\ldots$ are oscillatory terms which are irrelevant for our considerations. Here $x$ denotes a continuous position and $\phi(x)$ a bosonic field.  We emphasize that a spatially varying $K$ is not equivalent to a spatially varying two-particle interaction. In the process of switching to the effective field theory, a spatially varying two-particle interaction manifests itself in additional nonquadratic terms in the Hamiltonian in addition to a spatially varying $K$~\cite{contact}. More details on the field theory for an individual lead and the (bosonic) scattering theory if two leads are coupled via a device are given in Appendix \ref{app_B}. The dc conductance of such a setup can be computed~\cite{PhysRevB.52.R5539} as
\begin{align}
G =&\lim_{\omega \rightarrow 0^+}\frac{- i\omega}{\pi} \int_{-\infty}^{\infty} \frac{dt}{2\pi} e^{i\omega t}\left\langle T[\phi(x_1,t)\phi(x_2,0)]\right\rangle \nonumber \\
=& K_{\mathrm{L}} [1+(R_{\mathrm{L}}+R^{*}_{\mathrm{L}})/2],
\label{bosonizationDC}
\end{align}
where we have chosen $x_1, x_2$ to be in the left region and $R_{\mathrm{L}}$ is defined in the Appendix \ref{app_B}. In the case of an adiabatic contact, which constitutes the device,  Eqs.~(\ref{maxtransmissionRS}) hold and the maximal conductance
\begin{align}\label{Keff}
G=K_{\mathrm{eff}}=\frac{2K_{\mathrm{L}}K_{\mathrm{R}}}{K_{\mathrm{L}}+K_{\mathrm{R}}}    
\end{align}
is reached~\cite{PhysRevB.52.R17040}. 

In Appendix~\ref{app_B}, we also compute the charge fluctuations using the field theory and bosonization. We show that Eq.~(\ref{cvscalinginf}) with $G$ given by Eq.~(\ref{bosonizationDC}) [and thus Eq.~(\ref{Keff}] in the perfect transmission limit) and $G_{\rm h} = K_{\mathrm{L}}$ indeed holds. However, as for noninteracting leads, the ultimate verification of Eq.~(\ref{cvscalinginf}) beyond the field theory limit, with bulk two-particle interactions, and devices of different types is provided by the numerical evidence presented in Sec.~\ref{interactingleads}.

\subsection{The conductance and the bipartite entanglement entropy}

In this section, we, in addition, discuss the relation between $G$ and the finite-size scaling of the bipartite entanglement entropy $S$. Given the reduced density matrix $\rho$ of a subsystem, $S$ is defined as~\cite{PhysRevLett.90.227902}
\begin{align} \label{Sdef}
S=-\operatorname{Tr}\rho\ln(\rho). 
\end{align}
The finite-size scaling of $S$ for systems with impurity interfaces was investigated earlier~\cite{Peschel_2005,andp,Peschel_2012}.  Similar to the charge fluctuations, the entanglement entropy between one lead and the rest of the system is expected to diverge logarithmically with system size $L$, once the system is gapless and the low-energy excitations have a linear dispersion relation
\begin{align}\label{etg}
S = \sum_i\frac{c_{\text{eff}}(T_i)}{6}\ln(L)+ \ldots. 
\end{align}
The sum is over all modes. For some models, the so-called effective central charge $c_{\text{eff}}$, is not only a mathematical coefficient but is a function of $T_i$, which in turn is related to physical quantities. For a single-channel free fermion system with a single bond defect (device), one obtains~\cite{andp}
\begin{align}
c_{\text{eff}}(G)= & -\left(\frac{6}{\pi^2}\right)  \left\{ \left(1-\sqrt{G}\right) \operatorname{Li}_2\left(\sqrt{G}\right) \right. \nonumber \\ & +  \left(1+\sqrt{G}\right) \operatorname{Li}_2\left(-\sqrt{G}\right)  + \ln\left(\sqrt{G}\right)\nonumber \\ & \;\;\;\;  \times  \left[ \left(1+\sqrt{G}\right) \ln\left(1+\sqrt{G}\right) \right.  \nonumber \\ & \;\;\;\; \left. \left. +\left(1-\sqrt{G}\right)\ln\left(1-\sqrt{G}\right)\right]\right\} .
\label{ceff}
\end{align}
In Sec.~\ref{noninteractingleads}, we provide numerical evidence that Eq.~\eqref{ceff} also holds for a model with an interacting device (but noninteracting leads).

\section{Numerical Methods}\label{numericalmethod}

In Secs.~\ref{IIIA} and ~\ref{IIIB}, we describe the numerical methods to extract the conductance from either static charge fluctuations or a time-dependent approach.  For both methods, MPSs are used to represent wave functions. The numerical costs of these methods and some other MPS-based methods are discussed in Sec.~\ref{IIIC}.

\subsection{The charge fluctuation fitting and the entanglement entropy}\label{IIIA}

The major numerical task for the charge fluctuation scaling approach is to obtain the ground state of the Hamiltonian describing the device connected to two finite leads both having OBCs at the ends opposite to the device. A key for obtaining a reliable estimate for $G$ from charge fluctuations data is to consider proper subleading corrections.

We first obtain the ground state of the lattice model of interest (device coupled to the two finite leads) using DMRG. The two leads have OBCs on the ends opposite to the device. This makes the DMRG algorithm particularly efficient~\cite{RevModPhys.77.259}. Fully interacting (leads and device) as well as locally (device) interacting systems can be treated on equal footing. We fix the number of total sites to be even. For models with a device defined on an even number of sites, the left and right leads have an equal number of sites. Otherwise, the size of the two leads differ by one site. We work with fixed particle numbers. The models with the Hamiltonians Eqs.~\eqref{H1} and \eqref{HIL} (see below) have particle-hole symmetry (PHS). For these, we consider half filling of the entire system, which, given PHS, implies that both, the leads and the device are half filled. For the other models, we consider different fillings close to half filling and select the ground state with the lowest energy. 

As charge conservation is implemented in the MPS procedure, the charge fluctuations can efficiently be evaluated using the charge-resolved entanglement spectra. From the MPS procedure, the bipartite reduced density matrix in diagonal form $\sum_{i} \lambda^2_i |v_i\rangle \langle v_i|$ can directly be obtained, with the vectors $|v_i\rangle$ in the reduced Hilbert space having a definite charge. Doing statistics with the corresponding charges leads to the charge fluctuations.

When extracting $G$ from Eq.~\eqref{cvscaling} for models with noninteracting leads, we further consider subleading corrections. Due to charge fluctuations with wave vector $2k_{\mathrm{F}}$, the subleading corrections do not converge as a series in $L$. However, a subseries with equal $2k_{\mathrm{F}} L \!\! \mod 2\pi$ is convergent. For $k_{\mathrm{F}}=\pi/2$ (half filling), there are two such subseries. We can fit $G$ from either of the series using the following ansatz:
\begin{align}\label{Gfitansatz}
\left\langle N_{p}^2 \right\rangle-\left\langle N_{p} \right\rangle^2 -\frac{G}{2\pi^2}\ln(L)=\tilde{b}_0+\frac{\tilde{b}_1\ln(L)+\tilde{b}_2}{L}.
\end{align}
This ansatz for the subleading terms is motivated by results for the homogeneous free chain~\cite{Song}.  We have the freedom to choose which part of the system belongs to the device and which to the leads. More precisely, we can select some of the lattice sites of the (homogeneous) leads, located close to the device, as being part of the device. We exploit this flexibility of the partitioning and select a situation in which the fitting of Eq.~(\ref{Gfitansatz}) to the finite-size charge fluctuation data leads to comparably small $|\tilde{b}_0|$, $|\tilde{b}_1|$, and $|\tilde{b}_2|$. This increases the accuracy of the fit. We fix an optimized choice for each model considered below. We also average the fitting results from the above-mentioned two subseries for the model with Hamiltonian Eq.~\eqref{HIL}. 

For models with interacting leads, we do not expect  Eq.~(\ref{Gfitansatz}) for the corrections to the leading $L$ dependence to hold. Power-law corrections, typical for Tomonaga-Luttinger liquids~\cite{PhysRevB.90.155129}, and partly associated with the scaling of the backscattering amplitude (inhomogeneity), might appear. For such models, we, thus, only fit the ansatz with $G$ and $\tilde{b}_0$ and average over two different partitions of the device and lead which results in a reduced  error. 

For the spinless model with noninteracting leads, we also fit the effective central charge from entanglement entropy data using the ansatz
\begin{align}\label{cefffitansatz}
S-\frac{c_{\text{eff}}}{6}\ln(L)=\bar{b}_0+\frac{\bar{b}_1}{L}.
\end{align}
The subleading terms are again motivated from results obtained in the homogeneous free chain~\cite{Song}. The entanglement entropy is computed employing Eq.~(\ref{Sdef}).

\subsection{Time-dependent simulations to obtain the conductance}\label{IIIB}
Our time-dependent simulations require us to determine the ground state but also the time evolution into a quasi-steady state with a Hamiltonian supplemented by a bias voltage. To derive the linear conductance, the bias voltage is taken as small as (numerically) possible.
We follow the method described in Ref.~\cite{Ueda}, except for the differences we will state explicitly in the following. More details of this method and its variants can be found in Refs.~\cite{Ueda,PhysRevB.79.235336,PhysRevLett.88.256403,WhiteFeiguin,schneider2006conductance, PhysRevLett.101.140601, PhysRevB.73.195304, andp2010, PhysRevLett.124.137701, kang2021twowire}.

The ground state of the entire Hamiltonian (device and finite-size leads), obtained by an initial DMRG run, is used as the initial state of a quench dynamics, performed by TEBD~\cite{TEBD}. At time $t=0$, all on-site energies in the left lead are raised by half of the bias voltage  $V_{\mathrm{bias}}/2>0$ while the on-site energies in the right lead are lowered by  $-V_{\mathrm{bias}}/2$ and the time evolution is performed by the original Hamiltonian complemented by this bias voltage. As the time step of the evolution, we use 0.05 for the spinless models with the Hamiltonians Eqs.~\eqref{H1} and~\eqref{HIL} and 0.025 for the spinful model with the Hamiltonian Eq.~\eqref{H2}.  We turn on the bias suddenly, unlike in Ref.~\cite{Ueda}. Once the length $L$ of the leads is sufficiently large and after some transient time $t_{\mathrm{tr}}$, a quasi-steady state is reached---meaning for a long time interval the current $I(x,t)$ computed takes the approximately constant value $I_{\rm st}$. Note that, if arbitrarily large times---requiring arbitrarily large systems---could be reached, the steady-state value of the current would be independent of the position (bond) $x$ at which the current is computed. For finite times the position, however, matters; see below.   

To improve the quality of the estimate for the conductance $\lim_{V_{\mathrm{bias}} \to 0} I_{\rm st}/V_{\mathrm{bias}}$  obtained this way, some data analysis tricks turned out to be useful in the past~\cite{schneider2006conductance, Ueda}. We summarize those we used. 

The quasi-steady $I(x, t)$ is oscillatory with amplitude $\sim 1/L$  and period $t_{\rm p} \sim 1/V_{\mathrm{bias}}$. The steady-state value $I_{\rm st}$ is estimated by the peak-dip average of $I(x, t)$ in one period for our spinless models. We use this scheme as it works better than an average over the  full period in examples of free Hamiltonians we can easily test. This scheme requires being able to simulate the dynamics up to approximately $t_{\mathrm{tr}}+t_{\rm p}$.  For the spinful model and small $V_{\mathrm{bias}}$,  e.g., $\sim 0.025$, we can  only simulate up to $t_{\mathrm{tr}}$ but not up to $t_{\mathrm{tr}}+t_{\rm p}$, due to the increased computational costs. Fortunately, for the model with Hamiltonian Eq.~\eqref{H2},  $I_{\rm st}$ can instead be estimated by some averaging over the position (bond) at which the current is computed, which gives an approximate constant current for a time interval $\gtrapprox t_{\rm tr}$~\cite{Ueda}. However, this leads to larger errors than the peak-dip average used for the spinless models. For our QPC model with the Hamiltonian Eq.~\eqref{HQPC} and the parameters we consider, $t_{\mathrm{tr}}$ is already close to our simulation time limit and no obvious spatial averaging scheme exists. This prevents us from obtaining time-dependent data.

To evaluate the finite-size corrections, we also performed time-dependent computations for different $L$. We observe that the corrections can be large for the model with the Hamiltonian Eq.~\eqref{H2}. We thus use an extrapolation assuming a $1/L$ finite-size correction for $I_{\rm st}$. 

It turned out that by fixing the bias voltage to a value of the order $10^{-2}$ we do not introduce a considerable error to our estimate of the steady-state current. We thus refrain from extrapolation in $V_{\rm bias}$. We, however, note that this fixed scale sets another low-energy cutoff, which, in combination with the energy-level spacing $\sim 1/L$, might prevent us from reaching the asymptotic low-energy limit for the models with interacting leads; see Sec.~\ref{interactingleads} for more on this.

\subsection{Computational costs}\label{IIIC} 

Finally, we discuss the computational cost of calculating the conductance using MPS algorithms in more detail. We argue in favor of our proposed approach from the perspective of entanglement.  Note that the entanglement of an MPS is bound by the logarithm of its bond dimension~\cite{SCHOLLWOCK201196}. A linearly increasing entanglement indicates that a simulation with a desired fixed accuracy possibly needs exponentially growing resources. 
For example, using a local basis, the entanglement scales linearly with the number of internal degrees of freedom of the fermions (1 or 2 in our examples), making it more difficult to treat models with spin. 
Furthermore, the size of the device matters as increasing the device size is expected to require a proportional increase in the lead size. 
Fitting the conductance from the charge fluctuations only needs the ground-state data of the simple OBC geometry with a logarithmic dependence of the entanglement on the system size described by Eq.~\eqref{etg}. We argue that this entanglement is generically smaller as compared to the one to deal with in other approaches. 

The zero-bias conductance can be obtained directly from considering the finite bias current and taking the limit $V_{\rm bias} \to 0$.  The time-dependent approach we use to compute the steady-state current is known to have an entanglement which grows linearly in time for a fixed position at which the current is determined~\cite{PhysRevLett.124.137701}. Based on the reasonable assumption that the transient time is proportional to the device size, the computational cost to reach the steady state is expected to be exponentially large in the device size. For accurate finite bias steady-state data of free fermions, an entanglement volume law was demonstrated~\cite{fraenkel2021entanglement}.  The entanglement barrier is mitigated but not eliminated by a recent work~\cite{PhysRevLett.124.137701} using a mixed basis. To access the low-bias voltage limit, higher accuracy is needed to resolve smaller currents, while the entanglement growth rate is expected to be decreasing with decreasing voltage.

Another approach to the conductance employs linear response theory and requires the computation of the current-current correlation function Eq.~\eqref{gkubo}. A straightforward way to obtain this is using real-time simulation and Fourier transformation. The time-evolution problem is a local quench~\cite{impurityMPS}, thus a volume law barrier is not expected to occur. However, considering test cases, we convinced ourselves that quenched states can have considerably larger entanglement than ground states (not shown). This is most likely one of the reasons why algorithms employed in the past use frequency-space methods~\cite{EPL,PhysRevB.75.241103,PhysRevB.96.195111,bischoff2019density}. Those methods need to target additional states in addition to the ground state. To obtain those requires considerably larger bond dimensions; the problem is mitigated by generalizing the target states~\cite{EPL,PhysRevB.96.195111}. 

In comparison to dynamical approaches to obtaining the conductance, static equilibrium methods only require ground states.  At least two alternative approaches to our charge fluctuation scaling exploiting this were suggested in the past. In the first, one is working with a flux-induced persistent current~\cite{PhysRevB.64.155319,PhysRevB.67.193303}. In the second, a static correlation function is computed~\cite{PhysRevLett.105.226803,PhysRevB.85.045120, PhysRevB.86.075451}. These methods were barely used for more complex models than spinless fermions and a device of the complexity of a quantum dot. One hurdle is that geometries with a loop are used, which roughly doubles the entanglement for DMRG compared to the simplest open chain geometry we use. Our charge fluctuation scaling is close in spirit to the static correlation function approach.  One key improvement of our method is that the information on the  Fermi velocities is not needed, unlike in the scheme of Ref.~\cite{PhysRevB.86.075451} and its variants~\cite{kang2021twowire}. The Fermi velocity is not known analytically for generic interacting models. Furthermore, we have a fitting scheme incorporating the subleading corrections, which significantly improves the accuracy.

\section{Results for microscopic models}\label{sec2}

\subsection{Interacting devices connected to noninteracting leads}
\label{noninteractingleads}

In this section, we present our numerical results for the conductance obtained by the costly time-dependent approach as well as the efficient charge fluctuation scaling. A variety of setups of increasing complexity are considered. For all cases in which the two methods are applicable, we find agreement within the error bounds of the approaches. For the model with an interacting QPC, the time-dependent approach can no longer be used. However, internal consistency checks and a comparison to approximate results give us confidence that charge fluctuation scaling can also be employed in this case.    

\subsubsection{A minimal model of an interacting device coupled to noninteracting leads: spinless fermions}\label{doubledotsplinless}

We first consider a double quantum dot as the interacting device which is coupled to two leads. The fermions are assumed to be spinless. The Hamiltonian is given by
\begin{align}
H_{\mathrm{DD}}=&-\sum_{x \neq 0, \pm 1}c^{\dagger}_x c_{x+1}-\sum_{x=0, \pm 1}t_{\mathrm{c}}c^{\dagger}_x c_{x+1}  + \mbox{H.c.} \nonumber \\ &+V \left(n_0-\frac{1}{2}\right) \left(n_1-\frac{1}{2}\right),   
\label{H1}
\end{align}
where $n_x=c^{\dagger}_xc_x$. The hopping in the leads is set to one.  The hopping between the two sites defining the double-dot as well as the coupling between the double-dot and the leads is set to $t_{\mathrm{c}} < 1$. The local nearest-neighbor two-particle interaction is denoted by $V$.  Due to the shift of the density by 1/2, for vanishing Fermi energy the double dot is half filled. We compute the open boundary ground state by DMRG with total system sizes $500$, $600$, $700$, and $800$ lattice sites~\footnote{We obtain the charge fluctuation data with the cut of the left lead and the device located between site -1 and 0, which, in fact is already inside the device.}. From the bipartite reduced density matrix (see Sec.~\ref{numericalmethod}), we compute the charge fluctuations and extract $G$ by fitting Eq.~(\ref{Gfitansatz}). We also fit the ansatz Eq.~(\ref{cefffitansatz}) to the entanglement entropy data and extract the effective central charge $c_{\mathrm{eff}}$. Within the time-dependent approach (TEBD) an estimation for $G$ is obtained using the parameters $V_{\mathrm{bias}}=0.05$ and a total of $800$ sites.

We first consider the noninteracting limit $V=0$. The exact value of $G$ can be computed analytically employing single-particle scattering theory and is given by $G=4/(t_{\mathrm{c}}+1/t_{\mathrm{c}})^2$. For $V=0$, the errors resulting from the charge fluctuation fitting, the time-dependent approach, and the MPS approximation can be estimated separately.  In addition to the MPS, we can use exact diagonalization data (noninteracting system) for both the charge fluctuation fitting and the time-dependent approach. In exact diagonalization, the error due to the finite bond dimension is avoided.  Besides this, the errors of the time-dependent approach result from the finite bias $V_{\mathrm{bias}}=0.05$ and the finite system size (here $800$ sites). For $t_{\mathrm{c}}=0.6$, these absolute errors are of the order of $10^{-4}$. For this $t_{\mathrm{c}}$, the conductance obtained by the charge fluctuation fitting has an error $\sim 10^{-5}$. For both DMRG and TEBD, the finite bond dimension ($\chi=800$) errors are of the order $10^{-4}$. 

\begin{figure}[t!]
\includegraphics[width=\columnwidth]{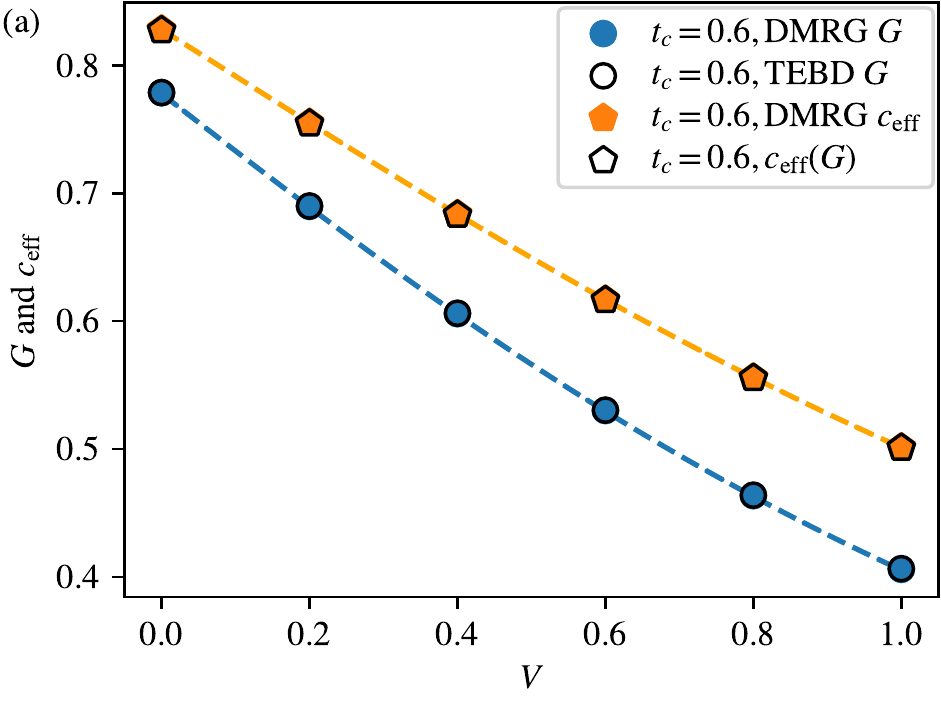}
\\[0.25cm]
\includegraphics[width=\columnwidth]{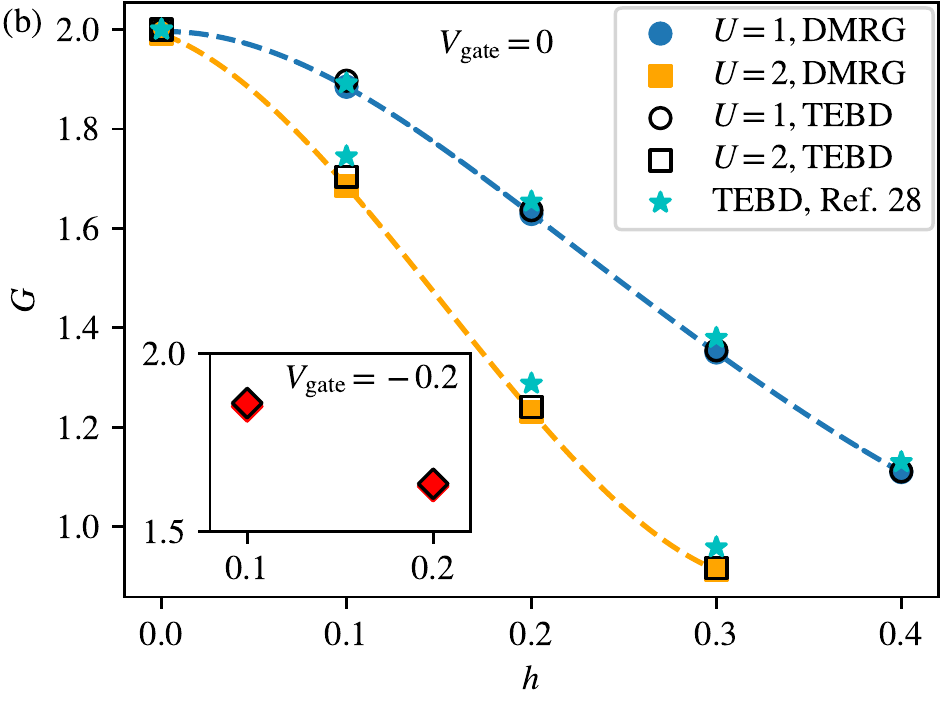}
\caption{(a) The  conductance $G$ and the effective central charge $c_{\rm eff}$ for the  double quantum dot model of spinless fermions, with $t_{\mathrm{c}}=0.6$ as a function of the two-particle interaction $V$. The filled symbols for $G$ are fitted from the DMRG data of the charge fluctuations with total system size 500, 600, 700, and 800.  The open symbols for $G$ show our TEBD results (time-dependent approach). The open symbols for $c_{\mathrm{eff}}(G)$ are computed by inserting the $G$ from the charge fluctuations in Eq.~\eqref{ceff}, while the filled ones are obtained directly from the entanglement entropy computed by DMRG. (b) The conductance $G$ of the single-impurity Anderson model with $t_{\mathrm{c}}=0.6$ as a function of the Zeeman field for fixed $U$ and $V_{\rm gate}$. Different $V_{\rm gate}$ are considered in the main plot and the inset. The filled symbols are fitted from the DMRG data of the charge fluctuations with total system size 136, 144, 152, and 160. Stars denote TEBD results from Ref.~\onlinecite{Ueda}, working with a system size 64 sites in total. In contrast, we here consider the $1/L$ correction within the  TEBD approach employing  system sizes of  128 and 256 sites. Our extrapolated data for $G$ are shown as open symbols. The dashed lines in (a) and (b) are guides to the eye obtained using cubic spline interpolation of the DMRG data [$G$ or $c_{\mathrm{eff}}(G)$].}
\label{fig:QD}
\end{figure}

We next investigate linear transport through an interacting double dot with $V>0$. We fix $t_{\mathrm{c}}=0.6$ and consider a varying amplitude of the two-particle interaction. In Fig.~\ref{fig:QD} (a), we plot the results for the conductance from the proposed charge fluctuation fitting procedure as filled circles. The conductance estimated from the time-dependent method is plotted as open circles. The latter show an agreement with the filled circles up to an average absolute deviation of $3\times 10^{-4}$ and a maximal absolute deviation of $10^{-3}$.  All this provides strong evidence that the conjectured relation between the finite-size scaling of the charge fluctuations and the conductance Eq.~\eqref{cvscaling} also holds for the interacting double dot. 

To examine the relation Eq.~\eqref{ceff} between the effective central charge and the conductance, we evaluate $c_{\mathrm{eff}}(G)$ using $G$ from the charge fluctuation fitting. In Fig.~\ref{fig:QD} (a), $c_{\mathrm{eff}}(G)$ is plotted as open pentagons, showing agreement with the filled pentagons, obtained directly from entanglement entropy fitting, up to $4\times 10^{-4}$ on average and a maximal absolute deviation of $6\times 10^{-4}$. This indicates that Eq.~\eqref{ceff} is applicable for the interacting double dot as well.

\subsubsection{A minimal model of an interacting device coupled to noninteracting leads: spinful fermions}

As another minimal model of an interacting device---this time concentrating on spinful fermions---coupled to noninteracting leads, we study the single-impurity Anderson model 
\begin{align}
H_{\mathrm{SIAM}}=&- \sum_{x \neq -1,0 \atop \sigma}  c^{\dagger}_{x,\sigma}c_{x+1,\sigma}- t_{\mathrm{c}} \sum_{x =-1,0 \atop \sigma}   c^{\dagger}_{x,\sigma}c_{x+1,\sigma} + \mbox{H.c.} \nonumber\\ \label{H2}
 &+U \left(n_{0, \uparrow}-\frac{1}{2}\right) \left(n_{0,\downarrow}-\frac{1}{2}\right)  \\
& + V_{\rm gate} \left( n_{0, \uparrow}+n_{0,\downarrow} \right) + h \left( n_{0, \uparrow}-n_{0,\downarrow} \right) \nonumber.    
\end{align}
Here $V_{\rm gate}$ denotes a gate voltage shifting the up- and down-spin dot levels and $t_{\rm c}$ has the same role as in the Hamiltonian Eq.~(\ref{H1}). The degeneracy of the two levels is lifted by a Zeeman field $h$. Due to the shift of the density by 1/2 in the two-particle term, at vanishing Fermi energy both levels are half filled for $V_{\rm gate}=0$ and $h=0$. For this configuration of single-particle parameters, a transport resonance $G=2$ occurs for any $U$~\cite{NgLee}. 

In Ref.~\onlinecite{Ueda} the conductance $G$ of this model was computed using TEBD for $V_{\rm gate}=0$, without considering finite-size corrections. Their data for the conductance as a function of the Zeeman field for two different interactions $U$ and $V_{\rm gate}=0$, using a total of 64 sites, are shown as the stars in the main panel of Fig.~\ref{fig:QD} (b). Our TEBD results---open symbols in Fig.~\ref{fig:QD} (b)---are based on data with 128 and 256 lattice sites in total as well as $1/L$ extrapolation. There is a clear discrepancy between the stars and our extrapolated data. The charge fluctuation fitting results of $G$ are shown as filled symbols. For this approach, we use data with system sizes of up to 160 sites. For $h=0$ and $V_{\rm gate}=0$, the deviation of the conductance obtained by the charge fluctuation fitting from the exact result $G=2$ is around $0.2\%$ for $U=1$ and  $0.4\%$ for $U=2$. For all the other parameters, the charge fluctuation fitting and TEBD results agree up to a relative difference of $0.5\%$ on average; the largest individual deviation is around $1\%$.  The TEBD data for the present model are less accurate than the others shown in this paper due to comparably large finite-size corrections. However, overall also this example gives us confidence that the charge fluctuation fitting can be used to extract the conductance and that it is a more efficient way to do so. 

\subsubsection{A spinful, interacting quantum point contact connected to noninteracting leads}

\begin{figure}[t]
\includegraphics[width=\columnwidth]{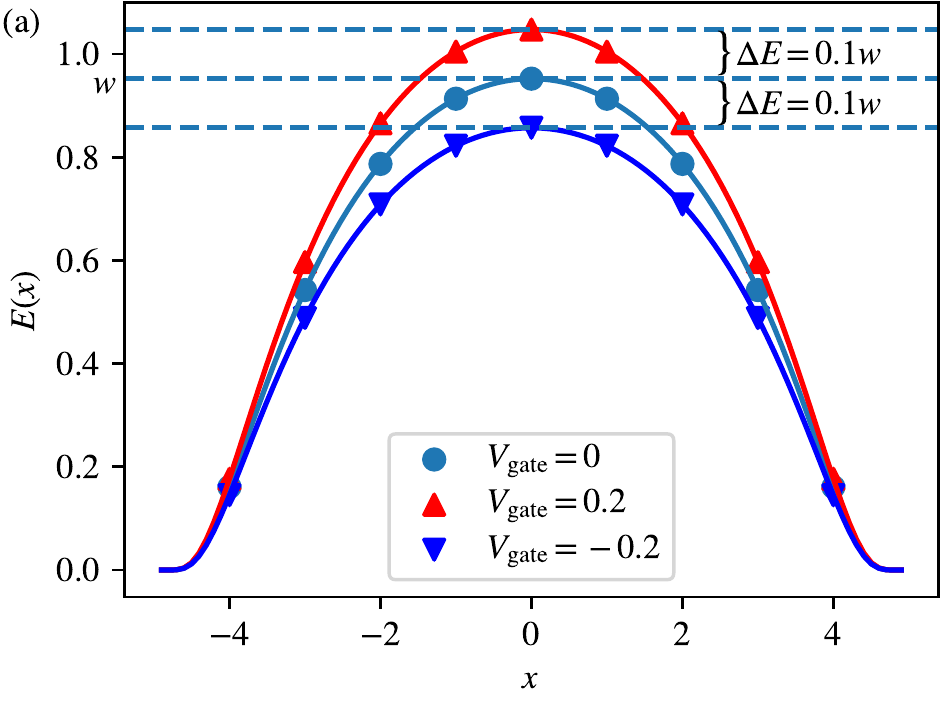}
\\[0.25cm]
\includegraphics[width=\columnwidth]{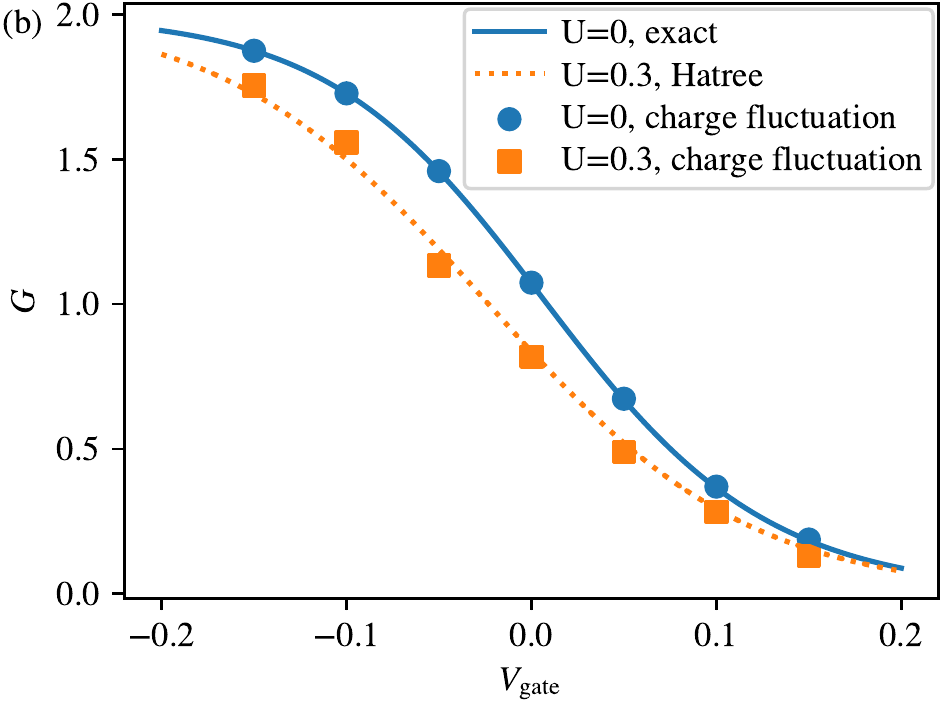}
\caption{(a) The spatial profile of the QPC potential for $W=5, b=0.55$. In the leads, the potential vanishes. The dashed line denotes the ``local bandwidth" and is tangential to the curve with $V_{\rm gate}=0$. This gate voltage corresponds to $G\approx 1$. (b) The conductance $G$ of the QPC model as a function of gate voltage $V_{\rm gate}$. The filled symbols are obtained by charge fluctuation scaling.  The solid line shows the ``numerically exact" conductance for  $U=0$. The dotted lines indicates the conductance obtained by the Hartree approximation for  $U=0.3$. }\label{fig:QPC}
\end{figure}

In this section, we illustrate that estimating the zero-temperature linear conductance from charge fluctuation scaling works for larger devices, whereas our time-evolution method is not viable due to its computational costs. We consider the model of a QPC introduced in Ref.~\onlinecite{Bauer2013}. It was employed to explain the 0.7 anomaly~\cite{PhysRevLett.77.135,PhysRevLett.88.226805} and is given by the Hamiltonian 
(in self-explaining notation) 
\begin{align}\label{HQPC}
H_{\mathrm{QPC}}=& - \!\!\!\!\!\! \sum_{x \in \mathrm{L}, \mathrm{R} \atop \sigma}  \!\!\!\!  c^{\dagger}_{x,\sigma}  c_{x+1,\sigma}- \!\!\!\! \sum_{x \in \mathrm{QPC} \atop \sigma} \!\!\!\!  t(x)c^{\dagger}_{x,\sigma}c_{x+1,\sigma} + \mbox{H.c.}  \nonumber \\ & +  \!\!\!\!\!\! \sum_{x \in \mathrm{QPC}}  \!\!\!\! \left[  U(x) n_{x, \uparrow}n_{x,\downarrow}+  E(x)(n_{x, \uparrow}+n_{x,\downarrow}) \right].    
\end{align}
The QPC region contains an odd number of sites $2W-1$. The  on-site potential in this region is $E(x)=\frac{V_{\rm gate}+2}{1+2b}e^{-(x/W)^2/[1-(x/W)^2]}$, can be controlled by a gate voltage $V_{\rm gate}$, and is shown in Fig.~\ref{fig:QPC} (a) for $W=5, \ b=0.55$. When $V_{\rm gate}=0$, the two-particle interaction $U(x)$ is set to zero, and the potential is chosen to match the effective, noninteracting, local bandwidth $w=2/(1+2b)$ (requiring  attenuation of the hopping $t(x)$ towards the QPC center), the QPC conductance is $G\approx 1$ ($G \rightarrow 1$ for $W \to \infty$).  More specifically, we use $t(x)=1-\frac{b}{2}[E(x)+E(x+1)]$. The local (on-site) interaction $U(x)$ decays from $U$ at the central site ($x=0$) towards  $0$ in the leads starting at sites $x= \pm W$. We employ $U(x)=U e^{-(x/W)^6/[1-(x/W)^2]}$.

We here focus on  $W=5, \ b=0.55$. On the one hand, our time-dependent simulation with TEBD or, alternatively, TDVP can only reach times comparable to the transient time. There is furthermore no obvious way for an averaging procedure (see above), which would allow us to extract reliable data for the steady-state current. Thus, we cannot obtain an estimation of the conductance from  a time-dependent approach for interacting QPCs.  On the other hand, we can estimate $G$ for $U=0$, and  $U=0.3$, using charge fluctuation data each with two sets of total system size $376,\ldots,400$ and $374,\ldots,398$. In Fig.~\ref{fig:QPC} (b), we plot the results as a function of the gate voltage as filled circles and squares, respectively. Error bars estimated by the difference of the extrapolation obtained from the two sets of system sizes are smaller than the symbol size.
In the noninteracting limit with $U=0$,  we plot the ``numerically exact" conductance obtained from a Green's function method~\cite{Oguri} as a solid line. The agreement between circles and the solid line is up to an absolute deviation $\sim 10^{-3}$. In Appendix \ref{app_C}, we compare the results for larger $W$ while keeping the total system size fixed. For the interacting case with $U=0.3$, we provide the conductance obtained from the Hartree approximation (the Fock term vanishes) as the dotted line in Fig.~\ref{fig:QPC} (b) for comparison. The consistency of the charge fluctuation fitting and the Hartree result gives us confidence that the charge fluctuation fitting can also be employed for an interacting QPC but also shows that the Hartree approximation seems to work well in the present setup and for moderate two-particle interactions. This indicates that, when studying the present model in the context of the 0.7 anomaly, the use of the Hartree approximation is sufficient and no other more elaborate approximate approach, such as, e.g., the truncated functional RG~\cite{Bauer2013}, is required.

\subsection{Systems with interacting leads}\label{interactingleads}

In this section, we consider the case that the leads are one-dimensional interacting metals. For simplicity, we restrict our attention to spinless fermions.  We consider the model with the Hamiltonian
\begin{align}\label{HIL}
H_{\mathrm{IL}}=&\sum_{x}\left[ \phantom{\frac{1}{2}} \!\!\!\!\!\! -t(x) \left( c^{\dagger}_xc_{x+1}+ \mbox{H.c.} \right) \right.\nonumber \\
& + \left.  V(x) \left(n_{x}-\frac{1}{2}\right)\left(n_{x+1}-\frac{1}{2}\right)\right].  
\end{align}
The nearest-neighbor hopping $t(x)$ and the nearest-neighbor two-particle interaction $V(x)$ take constant values in the left and right leads; the bulk interaction in these might differ ($t=1$, $V_{\mathrm{L}},\ V_{\mathrm{R}}$). The spatial region in which $t$ and $V$ depend on position defines the device (see Fig.~\ref{fig:models}). As discussed in Sec.~\ref{sec1}, in the low-energy limit those leads are described by the Tomonaga-Luttinger liquid theory, and characterized by a Tomonaga-Luttinger parameter $K$. The parameter $K$ can be extracted from static correlations. For a model with homogeneous nearest-neighbor hopping $t=1$ and homogeneous nearest-neighbor interaction $V$ at half filling, one finds~\cite{Haldane}
\begin{align}\label{xxzK}
K=\frac{\pi}{2[\pi-\arccos(V/2)]} .
\end{align}

\subsubsection{A smooth contact}\label{interactingleads_smooth}

\begin{figure}[tbp]
\includegraphics[width=\columnwidth]{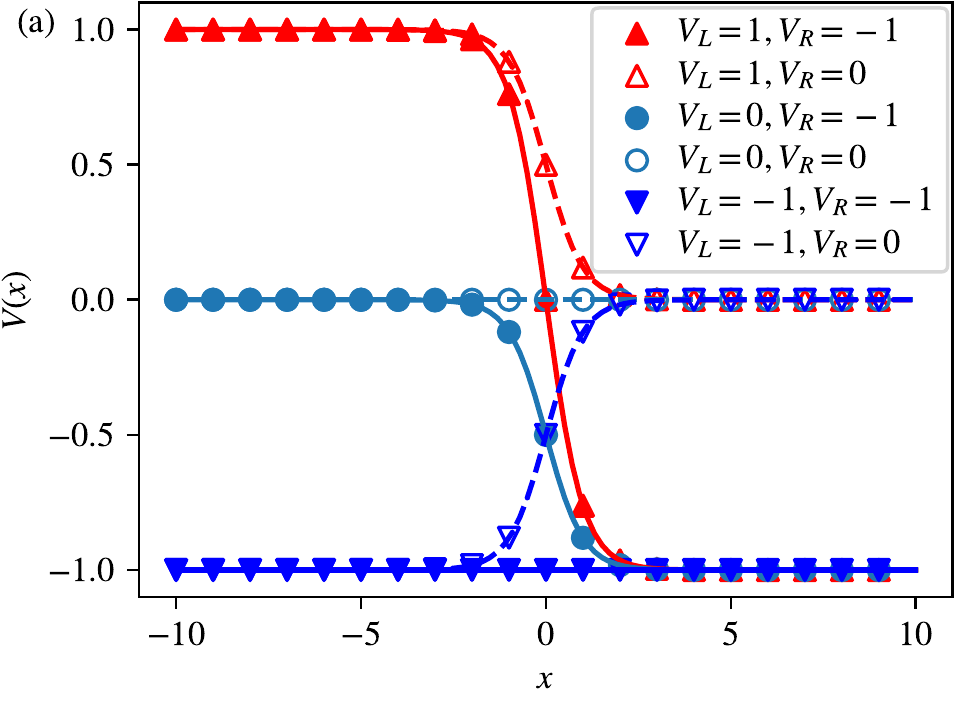}
\\[0.25cm]
\includegraphics[width=\columnwidth]{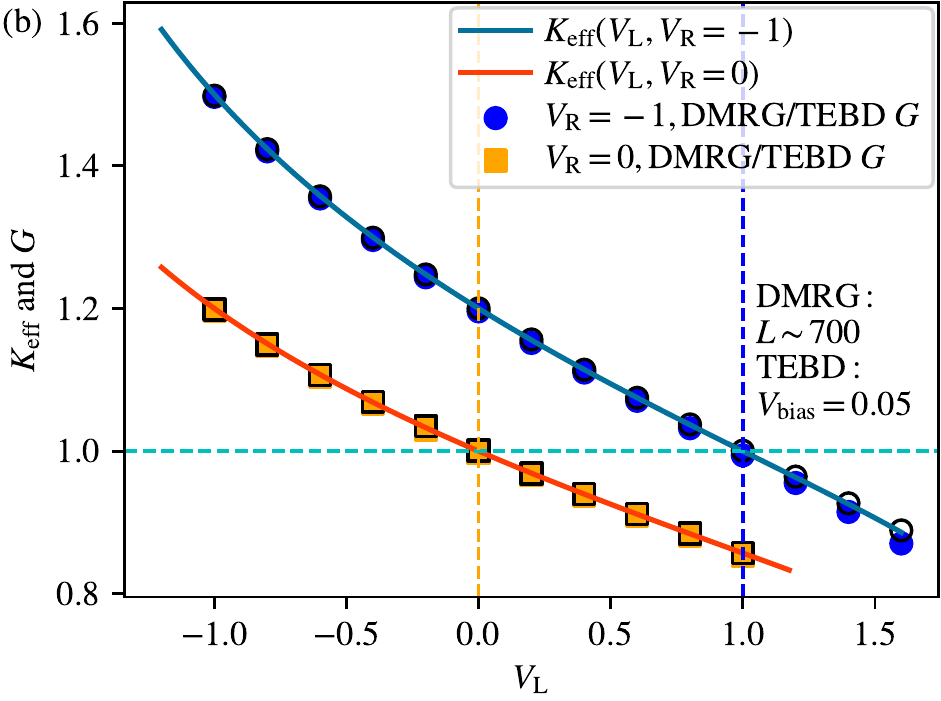} 
\caption{(a) Spatial profiles of the nearest-neighbor two-particle interaction described by  Eq.~\eqref{Vtanh} with $L_{\mathrm{c}}=1$. We study two families of profiles with $V_{\mathrm{R}}$ fixed at $-1$ (solid lines and filled symbols) as well as $0$ (dashed lines and open symbols). (b) The conductance as a function of $V_{\mathrm{L}}$ for two fixed $V_{\mathrm{R}}$. The charge fluctuation fitting results are shown as filled symbols and the data obtained from the time-dependent approach as open ones. For reference, the theoretical maximal conductance fixed point value $K_{\rm eff}$, computed from the two $K_{\mathrm{L},\mathrm{R}}$ is shown as solid lines. The charge fluctuation fitting was achieved with systems of sizes 500, 600, and 700. }\label{fig:xxz}
\end{figure}

In Sec.~\ref{sec1}, we discussed the fixed points of maximal and zero conductance obtained in setups in which the interaction takes different bulk values in the two leads. Here we firstly focus on numerically verifying the relation Eq.~\eqref{cvscaling} for the maximal conductance limit and consider $t(x)=1$. We choose a function $V(x)$ smoothly interpolating from left lead bulk interaction $V_{\mathrm{L}}$ to the right one $V_{\mathrm{R}}$. In particular, we consider
\begin{align}\label{Vtanh}
V(x)=\frac{V_{\mathrm{R}}-V_{\mathrm{L}}}{2}\tanh\left(  \frac{x}{L_{\mathrm{c}}} \right) + \frac{V_{\mathrm{L}}+V_{\mathrm{R}}}{2} ,
\end{align}
see Fig.~\ref{fig:xxz} (a). The length scale characterizing the crossover region, which here plays the role of the device,  is $L_{\mathrm{c}}$. We set $L_{\mathrm{c}}=1$ for the data plotted in Fig.~\ref{fig:xxz}.  

For definiteness, we investigate the part of the parameters space given by fixed $V_{\mathrm{R}}=-1$ or $0$ and a varying $V_{\mathrm{L}}$. Recall that in the  maximal conductance limit (adiabatic contact), $G=K_{\mathrm{eff}}=2K_{\mathrm{L}} K_{\mathrm{R}}/(K_{\mathrm{L}}+K_{\mathrm{R}})$, with $K_{\mathrm{L}}$ ($K_{\mathrm{R}}$) computed from $V_{\mathrm{L}}$ ($V_{\mathrm{R}}$) using Eq.~\eqref{xxzK}. In Fig.~\ref{fig:xxz} (b), we plot $G$, computed from the two $K_{\mathrm{L},\mathrm{R}}$, as solid lines for reference. The charge fluctuation scaling data for $G$, obtained for system sizes of up to 700 sites, are shown as filled symbols. The results of the time-dependent simulation with $V_{\mathrm{bias}}=0.05$ are shown as open symbols. All three data sets agree nicely for  $V_{\mathrm{R}}=-1$ and $0$. For  $V_{\mathrm{R}}=-1$, small but noticeable deviations can only be found for $V_{\mathrm{L}} \gtrapprox 1$. Overall this indicates that the relation between charge fluctuation scaling and the zero-temperature linear conductance also holds for interacting leads.

The influence of various inhomogeneities on the infrared fixed point of junctions of interacting quantum wires was investigated employing different approaches~\cite{PhysRevB.9.2911,PhysRevLett.68.1220,PhysRevB.46.15233, Meden_2008, contact, PhysRevB.86.075451, conductingfixedpoint1, conductingfixedpoint2, QMC2,bischoff2019density, kang2021twowire}. RG approaches within a tree-level approximation~\cite{PhysRevB.86.075451, conductingfixedpoint1, conductingfixedpoint2} show that, for models with different bulk interactions in the two leads, a bare finite backscattering will renormalize to zero, i.e., the system will approach the maximal conductance fixed point if $K_{\mathrm{eff}}>1$. For $K_{\mathrm{eff}}<1$, a finite backscattering will increase and the transmission vanishes in the infrared. In Fig.~\ref{fig:xxz} (b), we indicate  $K_{\mathrm{eff}}=1$ by a horizontal dashed line. This line intersects the two curves (with different $V_{\mathrm{R}}$) at one point per curve. Right of these two intersections, the tree-level analysis hints at a vanishing conductance. Based on this, we expect that the deviations of the charge fluctuations as well as the time-dependent data from $K_{\rm eff}$ for $V_{\mathrm{L}} \gtrapprox 1$ result from corrections to the adiabatic variation of the two-particle interaction from the left to the right bulk lead values. The maximal conductance fixed point is only stable if the spatial variation of the two-particle interaction is arbitrarily smooth. In a numerical computation for a lattice model, this limit can in practice never be reached. However, simultaneously we can also not reach asymptotically small energy scales, for which for a fixed smoothness the vanishing of the conductance due to the two-particle inhomogeneity should be observable~\footnote{Only one special data point in  Fig.~\ref{fig:xxz} (b) corresponds to a homogeneous interacting system. The one with $V=-0.5$.}.   In both our numerical approaches (charge fluctuation scaling and time-dependent simulation), this is prevented by the finite system size and the finite bond dimension, both implying a finite energy resolution. Increasing the smoothness of the interaction by increasing $L_{\mathrm{c}}$ in Eq.~\eqref{Vtanh} will reduce the effect of the nonadiabaticity down to some exceedingly small energy scale. 

\subsubsection{Transport resonances due to fine tuning}\label{interactingleads_tuning}

\begin{figure}[tbp]
\includegraphics[width=\columnwidth]{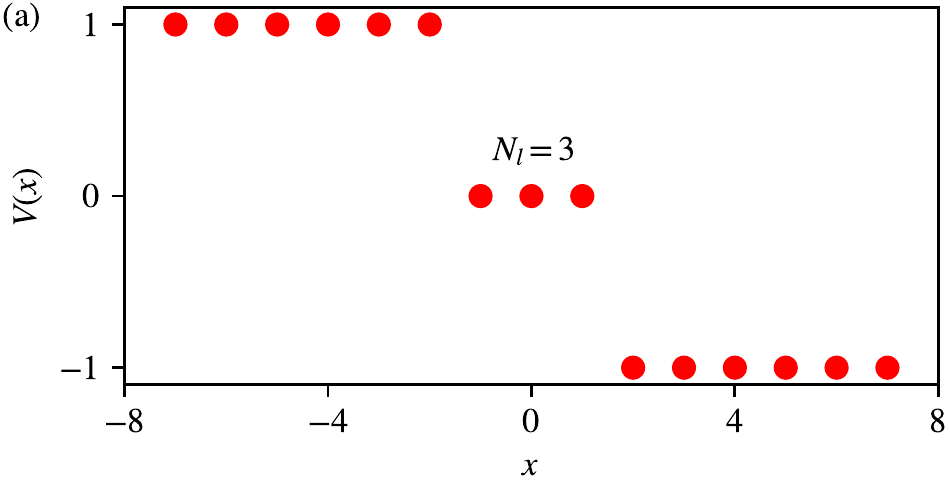}
\\[0.25cm]
\includegraphics[width=\columnwidth]{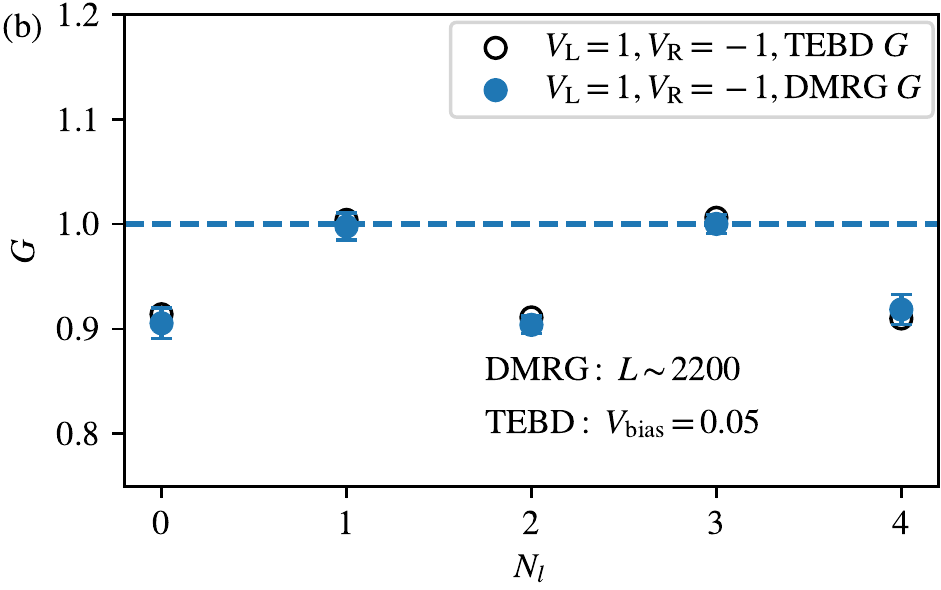} 
\caption{(a) Spatial profile of the inhomogeneous two-particle interaction described by  Eq.~\eqref{abruptV} with $N_l=3$ and $V_{\mathrm{L},\mathrm{R}}=\pm 1$. (b) The conductance as a function of $N_l$ for $V_{\mathrm{L},\mathrm{R}}=\pm 1$, leading to $K_{\rm eff}=1$, and $t(x)=1$. The charge fluctuation fitting results are shown as filled symbols and the data obtained from the time-dependent approach as open ones. The charge fluctuation fitting was achieved with system of sizes 2000, 2100, and 2200. The dashed horizontal line indicates the value for maximal conductance.} \label{fig:xxzabrupt} 
\end{figure}

While smooth interpolation of the two-particle interaction leads to weak backscattering, some fine-tuned nonsmooth configurations can also lead to weak or possibly even vanishing backscattering. We have already encountered one such example for the single-impurity Anderson model with $h=0$ and $V_{\rm gate}=0$, but arbitrary $U$ and $t_{\mathrm{c}}$ (transport resonance). Here, we illustrate a similar effect in the case of interacting leads. We consider the model with Hamiltonian Eq.~\eqref{HIL} and the special choice
\begin{align}\label{abruptV}
V(x)=\left\{
\begin{array}{ll}
    V_{\mathrm{L}},& x \leq x_{\mathrm{L}} \\
    0, & x_{\mathrm{L}}<x<x_{\mathrm{R}} ,\\
    V_{\mathrm{R}}, & x \geq x_{\mathrm{R}}
\end{array}
\right.
\end{align}
illustrated in Fig.~\ref{fig:xxzabrupt}(a) for $N_l=x_{\mathrm{R}}-x_{\mathrm{L}}-1=3$ and $V_{\mathrm{L}}=1$, $V_{\mathrm{R}}=-1$ . The abrupt change of the two-particle interaction can be expected to lead to backscattering. The case with $N_l=1$ was studied in Refs.~\onlinecite{PhysRevB.86.075451, kang2021twowire} where, in addition, different hoppings $t(x) \neq 1$ for $x_{\mathrm{L}}<x<x_{\mathrm{R}}$ were considered. For the case $t(x)=1$ for all $x$, no clear deviation from the maximal conductance was found for $K_{\mathrm{eff}}=1$. For some pairs of $K_{\mathrm{L}}$ and $K_{\mathrm{R}}$ as well as certain configurations with $t(x) \neq 1$, the effect of the abrupt change of the interaction was, in addition, found to be  weak~\cite{PhysRevB.86.075451}. Here, we report on a related even-odd effect for different $N_l$. 

In Fig.~\ref{fig:xxzabrupt} (b), we show that only for even $N_l$, there is a clear deviation from the maximal conductance $G=K_{\rm eff}=1$. This holds for both our methods to extract the conductance and the charge fluctuation scaling (filled symbols) as well as the time-dependent approach (open symbols). Again, both give results which agree within the respective error bounds. In Appendix~\ref{app_C}, we show additional data for the case $N_l=0$ and various values of the interaction.   We interpret the case for odd $N_l$ as a fine-tuning  phenomenon of weak or even absent backscattering (transport resonance).  

To argue in favor of this, we follow the tree-level analysis of Refs.~\onlinecite{conductingfixedpoint1, conductingfixedpoint2}. In this analysis, a quadratic backscaterring term in the effective field theory is constructed as $\lambda \psi^{\dagger}_L(0)\psi_R(0)+\mbox{H.c.}$, where $\psi_L$ and $\psi_R$ are the left- and right- moving fermion fields of the Tomonaga-Luttinger model~\cite{Giamarchi_book}. The parameter $\lambda$ is determined from the single-particle parameters (hopping and on-site potentials) as well as the inhomogeneous two-particle interaction. However, given a certain microscopic model, no closed form expression for this can be written for arbitrary interactions. Within field theory and the tree-level approximation, a bare $\lambda=0$ is associated with a fine-tuned maximal conductance fixed point with $G=K_{\mathrm{eff}}$. It was argued that a combination of the hopping and interaction inhomogeneities determines the bare effective $\lambda$ through the renormalized local Fermi velocity $\lambda \propto \sum_{x} (-1)^x v_{\mathrm{F}}(x)$~\cite{conductingfixedpoint1, conductingfixedpoint2}. Working at tree level, we also constrain the analysis to the weak coupling limit ($V_{\mathrm{L}}, V_{\mathrm{R}} \ll 1$). For a generic inhomogeneous model of the type Eq.~(\ref{HIL}) at half filling, to leading order in the two-particle interaction, the local renormalized velocity is given by $v_{\mathrm{F}}(x)=2[t(x)+V(x)/\pi]$.  In addition, for the lead Tomonaga-Luttinger liquid parameters, one obtains $K_{\mathrm{L},\mathrm{R}}=1-V_{\mathrm{L},\mathrm{R}}/\pi$ to leading order.  To this order, $K_{\mathrm{eff}}=1$ thus implies that $V_{\mathrm{L}}+V_{\mathrm{R}}=0$ holds. In fact, using the exact result (at half filling) Eq.~\eqref{xxzK}, we see that $K_{\mathrm{eff}}=1$ always implies $V_{\mathrm{L}}+V_{\mathrm{R}}=0$. With only the two-particle interaction being inhomogeneous for even $N_l$, $\lambda \propto V_{\mathrm{L}}-V_{\mathrm{R}}$ to leading order. For odd $N_l$, $\lambda \propto V_{\mathrm{L}}+V_{\mathrm{R}}$. Thus, to lowest order in the two-particle interaction, $\lambda$ vanishes for odd $N_l$. Curiously, this implication, derived in the weak interaction limit, survives for our data with relatively large coupling $V_{\mathrm{L}}=-1$ and $V_{\mathrm{R}}=1$. However, not observing a deviation from the maximal conductance in our computation does not imply that on exceedingly small energy scales such deviations might not occur (see above). To investigate this further would require other methods.\\  

\section{Conclusion}

We here suggested a relation between the charge fluctuations and the zero-temperature linear response conductance for a system with two homogeneous leads coupled to an inhomogeneous device. Both the device as well as the leads can be interacting. First, we used general considerations and insights from the noninteracting case to argue in favor of such a relation. Second, we numerically verified this relation for a variety of setups. For this, we compared the conductance extracted from charge fluctuation scaling, obtained by numerical DMRG  with the conductance computed from the finite bias current computed by numerical TEBD or by exact results, if available. For the charge fluctuation scaling, systems with open boundaries in the leads on the ends opposite to the device can be used. This brings about a significant computational advantage. For our models, the time-dependent method on average requires a ten times larger running time as compared to the charge fluctuation scaling. We, e.g., show that the charge fluctuation method leads to reliable data for a QPC model, which cannot be dealt with using our time-dependent method due to its computational cost. We also use the proposed method to study the backscattering effect due to interaction non-smoothness at the junction of two interacting leads. We find  that backscattering can be fine-tuned to a value so small, that it does not show on the comparably low energy scales accessible to us. While the data can be interpreted within a weak-coupling theory, our observations call for more investigations.

\begin{acknowledgments}
We thank Chang-Yu Hou, Sebastian Paeckel, Peter Schmitteckert, and Kang Yang for discussion. This  work  was  supported by the Deutsche Forschungsgemeinschaft (DFG, German  Research  Foundation)  via  RTG  1995. The numerical calculation is based on TeNPy packages~\cite{iDMRGpaper, 10.21468/SciPostPhysLectNotes.5}, and has been performed with computing resources granted by RWTH Aachen University under Project No.rwth0661.
\end{acknowledgments}
\appendix
\begin{widetext}

\section{} \label{app_A}

In this appendix, we consider one-dimensional noninteracting spinless fermions in the geometry shown in  Fig.~\ref{fig:cvblock} (a).  We analytically compute the charge fluctuations of a segment ($l$) in the left lead [cf.~Fig.~\ref{fig:cvblock} (a)] with length $l$. Fixing the right end of the segment at $x=l_0<0$, we concentrate on the leading term of the charge fluctuations for large $l$.

For noninteracting fermions, charge fluctuations are given by single particle correlation functions as follows
\begin{align}\label{wick}
\left\langle N_{l}^2 \right\rangle-\left\langle N_{l} \right\rangle^2&=\sum_{x,y \in l }\left[\langle n(x) n(y)\rangle- \langle n(x) \rangle \langle n(y)\rangle\right]\\
&=\sum_{x,y \in l } \langle c^{\dagger}(x) c(y) \rangle  \langle c(x) c^{\dagger}(y) \rangle \nonumber \\
&=\sum_{x \in l } \langle n(x) \rangle-\sum_{x,y \in l } \langle c^{\dagger}(x) c(y) \rangle^2.   \nonumber
\end{align}

The final goal is to compute the large $l$ expansion of $\left\langle N_{l}^2 \right\rangle-\left\langle N_{l} \right\rangle^2$. To do this, we first compute $\langle c^{\dagger}(x) c(y) \rangle$ for $x,y$ in the left lead, using the field operator solution [cf.~Eq.~\eqref{scat_state}] of 
\begin{align}
c(x)= \frac{1}{\sqrt{2\pi}}\int_{0}^{\pi} dk \left[c_{\mathrm{L}}(k)(e^{ikx}+r_{\mathrm{L}}(k)e^{-ikx})+c_{\mathrm{R}}(k)t_{\mathrm{R}}(k)e^{ikx}\right],
\end{align}
where $c_{\mathrm{L}}(k)$ ($c_{\mathrm{R}}(k)$) annihilates a left [right] scattering state with momentum $k>0$ and $t_{\mathrm{L},\mathrm{R}}(k)$ and $r_{\mathrm{L},\mathrm{R}}(k)$ are the transmission and reflection amplitude, respectively. 
The above solutions do not include possible bound states, as they can be neglected when considering the leading term of charge fluctuations.
In the following, we assume the ground state has a pair of Fermi points at $ \pm k_{\mathrm{F}}$ and that the states with momentum between them are filled.  The single-particle correlation function $\langle c^{\dagger}(x) c(y) \rangle$ within a lead can be decomposed to a Toeplitz part, depending on $x-y$ and a Hankel part, depending on $x+y$.  For the left lead, we find: 
\begin{align}\label{cdaggerc}
\langle c^{\dagger}(x) c(y) \rangle= &\frac{1}{2\pi} \int_{0}^{k_{\mathrm{F}}} dk \left[e^{ik(y-x)}+e^{ik(x-y)}|r_{\mathrm{L}}(k)|^2+e^{-ik(x+y)}r_{\mathrm{L}}(k)+e^{ik(x+y)}r^{*}_{\mathrm{L}}(k)+e^{ik(x-y)}|t_{\mathrm{R}}(k)|^2\right]\\
=& \frac{1}{2\pi} \int_{0}^{k_{\mathrm{F}}} dk \left[e^{ik(y-x)}+e^{ik(x-y)}+e^{-ik(x+y)}r_{\mathrm{L}}(k)+e^{ik(x+y)}r^{*}_{\mathrm{L}}(k)\right] \nonumber \\
=& \frac{\sin[k_{\mathrm{F}}(x-y)]}{\pi(x-y)}+ \frac{1}{2\pi} \int_{0}^{k_{\mathrm{F}}} dk \left[e^{-ik(x+y)}r_{\mathrm{L}}(k)+e^{ik(x+y)}r^{*}_{\mathrm{L}}(k)\right] \nonumber \\ \equiv  & G_{\mathrm{T}}(x-y)+G_{\mathrm{H}}(x+y). \nonumber
\end{align}
With Eq.~\eqref{cdaggerc}, Eq.~\eqref{wick} becomes
\begin{align}\label{cfsum}
\left\langle N_{l}^2 \right\rangle-\left\langle N_{l} \right\rangle^2=\sum_{x \in l } \left[\frac{k_{\mathrm{F}}}{\pi} + G_{\mathrm{H}}(2x)\right]-\sum_{x,y \in l} \left[G^2_{\mathrm{T}}(x-y)+ 2G_{\mathrm{T}}(x-y)G_{\mathrm{H}}(x+y)+G^2_{\mathrm{H}}(x+y)\right] .
\end{align}

Then we analyze the large $l$ expansion of Eq.~\eqref{cfsum} by analyzing separate groups of terms individually. We group the terms of Eq.~\eqref{cfsum} into two parts:
\begin{align}\label{variancedeomp}
\left\langle N_{l}^2 \right\rangle-\left\langle N_{l} \right\rangle^2=\left(\left\langle N_{l}^2 \right\rangle-\left\langle N_{l} \right\rangle^2\right)_{\mathrm{homogeneous}}+\left(\left\langle N_{l}^2 \right\rangle-\left\langle N_{l} \right\rangle^2\right)_{\mathrm{residual}},
\end{align}
where
\begin{align}\label{variancehom}
\left(\left\langle N_{l}^2 \right\rangle-\left\langle N_{l} \right\rangle^2\right)_{\mathrm{homogeneous}}=\sum_{x \in l } \frac{k_{\mathrm{F}}}{\pi} -\sum_{x,y \in l} G^2_{\mathrm{T}}(x-y),
\end{align}
and
\begin{align}\label{varianceres}
\left(\left\langle N_{l}^2 \right\rangle-\left\langle N_{l} \right\rangle^2\right)_{\mathrm{residual}}=\sum_{x \in l } G_{\mathrm{H}}(2x)-\sum_{x,y \in l} \left[2G_{\mathrm{T}}(x-y)G_{\mathrm{H}}(x+y)+G^2_{\mathrm{H}}(x+y)\right].
\end{align}
By definition, $\left(\left\langle N_{l}^2 \right\rangle-\left\langle N_{l} \right\rangle^2\right)_{\mathrm{homogeneous}}$ denotes the charge fluctuation in a segment of a homogeneous free fermion chain with a pair of Fermi points. The leading term in this case is known to be~\cite{Song} 
\begin{align}\label{variancehom}
\left(\left\langle N_{l}^2 \right\rangle-\left\langle N_{l} \right\rangle^2\right)_{\mathrm{homogeneous}}= \frac{1}{\pi^2}\ln(l)+\ldots,
\end{align}
where $ \ldots $ denotes subleading terms in the large $l$ limit.

The remaining task is to analyze the leading term of Eq.~\eqref{varianceres} for large $l$. It is helpful to first consider the leading term of $G_{\mathrm{T}}(x)$ and $G_{\mathrm{T}}(x)$ in the  large $|x|$ limit. A closed form expression of $G_{\mathrm{T}}(x)$ is known (see above) and we derive  the large $|x|$ expansion of  $G_{\mathrm{H}}(x)$ to be: 
\begin{align}~\label{GHleading}
G_{\mathrm{H}}(x)=\frac{|r_{\mathrm{L}}(k_{\mathrm{F}})|}{\pi}\frac{\sin\left[k_{\mathrm{F}} x+\delta(k_{\mathrm{F}})\right]}{x}+ \ldots,    
\end{align}
where 
\begin{align}
\delta(k_{\mathrm{F}})=\arg \big[r_{\mathrm{L}}(k_{\mathrm{F}})\big] . 
\end{align}
This can be shown as follows. Recalling the definition 
\begin{align}
G_{\mathrm{H}}(x)=\frac{1}{2\pi} \int_{0}^{k_{\mathrm{F}}} dk \left[e^{-ikx}r_{\mathrm{L}}(k)+e^{ikx}r^{*}_{\mathrm{L}}(k)\right] ,
\end{align}
we  integrate by parts and obtain
\begin{align}~\label{GHbypart}
G_{\mathrm{H}}(x)=&\frac{1}{2\pi} \frac{i}{x} \int_{0}^{k_{\mathrm{F}}} dk \left[(e^{-ikx})^{\prime} r_{\mathrm{L}}(k)-(e^{ikx})^{\prime}r^{*}_{\mathrm{L}}(k)\right]\\
=& \frac{1}{2\pi} \frac{i}{x} \left[e^{-ikx} r_{\mathrm{L}}(k)-e^{ikx}r^{*}_{\mathrm{L}}(k)\right] |_{k=0}^{k_{\mathrm{F}}} -\frac{1}{2\pi} \frac{i}{x} \int_{0}^{k_{\mathrm{F}}} dk \left[e^{-ikx} r^{\prime}_{\mathrm{L}}(k)-e^{ikx}r^{\prime*}_{\mathrm{L}}(k)\right] \nonumber \\ 
=& \frac{1}{2\pi} \frac{i}{x} \left[e^{-ik_{\mathrm{F}}x} r_{\mathrm{L}}(k_{\mathrm{F}})-e^{ik_{\mathrm{F}}x}r^{*}_{\mathrm{L}}(k_{\mathrm{F}})\right] -\frac{1}{2\pi} \frac{i}{x} \int_{0}^{k_{\mathrm{F}}} dk \left[e^{-ikx} r^{\prime}_{\mathrm{L}}(k)-e^{ikx}r^{\prime*}_{\mathrm{L}}(k)\right] \nonumber \\
=& \frac{1}{2\pi} \frac{i}{x} \left[e^{-ik_{\mathrm{F}}x} r_{\mathrm{L}}(k_{\mathrm{F}})-e^{ik_{\mathrm{F}}x}r^{*}_{\mathrm{L}}(k_{\mathrm{F}})\right] +O(1/x^2). \nonumber 
\end{align}
To obtain the second last line of Eq.~\eqref{GHbypart}, we have used that $r_{\mathrm{L}}(k=0)$ is  real [because the linear equations for solving $r_{\mathrm{L}}(k=0)$ and $t_{\mathrm{L}}(k=0)$ are real]. To obtain the last line of Eq.~\eqref{GHbypart}, we have used the assumption that $\int_{0}^{k_{\mathrm{F}}} dk \left[e^{-ikx} r^{\prime}_{\mathrm{L}}(k)-e^{ikx}r^{\prime*}_{\mathrm{L}}(k)\right]=O(1/x)$. A sufficient condition for this assumption to hold is that $r_{\mathrm{L}}(k)$ is  second-order derivable. This can be shown by another integration by parts 
\begin{align}\label{furtherintegatebypart}
\int_{0}^{k_{\mathrm{F}}} dk \left[e^{-ikx} r^{\prime}_{\mathrm{L}}(k)-e^{ikx}r^{\prime*}_{\mathrm{L}}(k)\right]=\frac{i}{x} \left[e^{-ikx} r^{\prime}_{\mathrm{L}}(k)-e^{ikx}r^{\prime*}_{\mathrm{L}}(k)\right] |_{k=0}^{k_{\mathrm{F}}}- \frac{i}{x} \int_{0}^{k_{\mathrm{F}}} dk \left[e^{-ikx} r^{\prime\prime}_{\mathrm{L}}(k)-e^{ikx}r^{\prime\prime*}_{\mathrm{L}}(k)\right].
\end{align}
The term in Eq.~\eqref{furtherintegatebypart} is $O(1/x)$ because the terms within square brackets are $O(1)$.
Thereby, we have obtained Eq.~\eqref{GHleading} from Eq.~\eqref{GHbypart}.

Now we can complete the analysis of the leading term of Eq.~\eqref{varianceres}. We argue that the leading term of Eq.~\eqref{varianceres} comes only from its third term, which contributes a leading logarithmic divergence in the large $l$ limit:
\begin{align}~\label{sumgh}
\sum_{x,y \in l}G^2_{\mathrm{H}}(x+y)=\frac{|r_{\mathrm{L}}(k_{\mathrm{F}})|^2}{2\pi^2}\ln(l)+ \ldots.    
\end{align}
Equation.~\eqref{sumgh} can be derived as follows. Using Eq.~\eqref{GHleading}, we obtain
\begin{align}\label{variancehankel}
\sum_{x,y \in l} G^2_{\mathrm{H}}(x+y) &=\frac{|r_{\mathrm{L}}(k_{\mathrm{F}})|^2}{\pi^2} \sum_{x,y \in l}\frac{\cos^2\left[k_{\mathrm{F}} (x+y)+\delta(k_{\mathrm{F}})\right]}{(x+y)^2}+ \ldots, \nonumber \\
& = \frac{|r_{\mathrm{L}}(k_{\mathrm{F}})|^2}{2\pi^2} \sum_{x,y \in l}\frac{1+\cos\left[2k_{\mathrm{F}} (x+y)+2\delta(k_{\mathrm{F}})\right]}{(x+y)^2}+ \ldots, \nonumber \\
& = \frac{|r_{\mathrm{L}}(k_{\mathrm{F}})|^2}{2\pi^2} \sum_{x,y \in l}\frac{1}{(x+y)^2}+ \ldots, \nonumber\\
& = \frac{|r_{\mathrm{L}}(k_{\mathrm{F}})|^2}{2\pi^2} \ln(l) + \ldots , 
\end{align}
where $ \ldots $ denotes subleading terms in the large $|x+y|$ or $l$ limit. 
The last line can be justified by recognizing that there are lower and upper bounds to the sum by the two integrals: $\int_{x,y=l_0-l-1}^{l_0} dx dy \frac{1}{(x+y)^2} <\sum_{x,y=l_0-l}^{l_0}\frac{1}{(x+y)^2}<\int_{x,y=l_0-l}^{l_0+1} dx dy \frac{1}{(x+y)^2}$ and  realizing that the two integrals both have a leading term $\ln(l)$. For the third line, we have neglected the term with oscillatory factor $\cos[2k_{\mathrm{F}} (x+y)+2\delta(k_{\mathrm{F}})]$, which does not contribute to the divergence; this is because summing within each period of  the oscillatory factor yields a series decaying faster than $\frac{1}{(x+y)^2}$ for further summation. Having analyzed the third term of Eq.~\eqref{varianceres}, second we show the other two terms only contribute subleading terms and thus can be neglected. The first term  of Eq.~\eqref{varianceres} [$\sum_{x \in l } G_{\mathrm{H}}(2x)$] does not contribute to the divergence  due to its oscillatory factor again.  For the second term,  we perform the summation by changing the  variables $r_{\rm s}=x+y$ and $r_{\rm d}=x-y$. 
\begin{align}
-\sum_{x,y \in l} 2G_{\mathrm{T}}(x-y)G_{\mathrm{H}}(x+y)=-\sum_{2x_{\mathrm{min}}\leq r_{\rm s} \leq 2x_{\mathrm{max}}} G_{\mathrm{H}}(r_{\rm s}) A(r_{\rm s},x_{\mathrm{min}}, x_{\mathrm{max}}) 
\end{align}
We expect that $A(r_{\rm s},x_{\mathrm{min}},x_{\mathrm{max}})$ can be approximated as a constant in the large $l$ and $|r_{\rm s}|$ limit, and the above summation converges also due to the oscillatory factor of $G_{\mathrm{H}}$.

Combining Eqs.~\eqref{variancedeomp},~\eqref{varianceres}, ~\eqref{variancehom}, and~\eqref{variancehankel} and remembering that $G=|t_{\mathrm{R}}(k_{\mathrm{F}})|^2=1-|r_{\mathrm{L}}(k_{\mathrm{F}})|^2$,  we obtain 
\begin{align}
\left\langle N_{l}^2 \right\rangle-\left\langle N_{l} \right\rangle^2=\frac{2-|r_{\mathrm{L}}({k_{\mathrm{F}}})|^2}{2\pi^2}\ln(l)+ \ldots =\frac{1+G}{2\pi^2}\ln(l)+ \ldots.
\end{align}

\end{widetext}

\section{} \label{app_B}

Here, we provide arguments for Eq.~\eqref{cvscalinginf} in the case of interacting leads. We work with an effective field theory and employ bosonization~\cite{Giamarchi_book,PhysRevB.52.R17040}.   Up to the contact to the device, the lead Hamiltonian is homogeneous. Thus, the effective theory of the leads after neglecting terms which are irrelevant in the RG sense is a noninteracting bosonic field theory~\cite{Giamarchi_book}. For example, in the left lead, the Lagrangian density is $\mathcal{L}=\frac{1}{2\pi K_{\mathrm{L}}}\left\{ -v_{\mathrm{L}} \left[\partial_{x}\phi(x,t)\right]^2+\frac{1}{v_{\mathrm{L}}}\left[\partial_{t}\phi(x,t)\right]^2\right\}$. Here $v_{\mathrm{L}}$ denotes the renormalized Fermi velocity of the left lead. This implies that the field $\phi(x,t)$, determining the density, obeys the equation of motion $v^2_{\mathrm{L}} \partial^2_x \phi(x,t)- \partial^2_t \phi(x,t) =0$, which has plane-wave solutions.  It also obeys the canonical commutation relation $[\phi(x,t), \Pi(x^{\prime},t)]=i \delta(x-x^{\prime})$, where $\Pi(x,t)=\frac{1}{\pi K_{\mathrm{L}} v_{\mathrm{L}}} \partial_{t}\phi(x,t)$ in the left lead.

Suppose that the scattering can be encoded by mixing of the $\pm k$ plane wave modes, we rewrite the field as
\begin{align}
\phi(x,t) = & \frac{\sqrt{K_p}}{2\sqrt{\pi}}\int_{0}^{+\infty} \frac{d k}{\sqrt{2|k|}}  \big[ u_{\mathrm{L},k}(x) a_k e^{-i\omega_{\mathrm{L},k} t} \\ & -u_{\mathrm{R},k}(x) b_k e^{-i\omega_{\mathrm{L},k} t} \big] 
 + \mbox{H.c.},  \nonumber
\end{align}
where $p= \mathrm{L}$ ($\mathrm{R}$) for the left (right) lead, $\omega_{\mathrm{L},k}=v_{\mathrm{L}} |k|$, and
\begin{align}
&u_{\mathrm{L},k}(x)= e^{ikx}+R_{\mathrm{L}}(k)e^{-ikx} ,x \in \text{left region}, \nonumber \\
&u_{\mathrm{R},k}(x)=S_{\mathrm{R}}(k)e^{-ikx},x \in \text{left region}, \nonumber \\
&u_{\mathrm{L},k}(x)= S_{\mathrm{L}}(k)e^{ikxv_{\mathrm{L}}/v_{\mathrm{R}}} ,x \in \text{right region}, \nonumber \\
&u_{\mathrm{R},k}(x)= e^{ikxv_{\mathrm{L}}/v_{\mathrm{R}}}+R_{\mathrm{R}}(k)e^{-ikxv_{\mathrm{L}}/v_{\mathrm{R}}} ,x \in \text{right region} .
\end{align}
Here $S_{\mathrm{L}}(k)$, $R_{\mathrm{L}}(k)$ [$S_{\mathrm{R}}(k)$, $R_{\mathrm{R}}(k)$] are coefficients of the solutions in the left (right) region. These coefficients are also related to the conductance and the static density correlations, but with a different relation [see Eq.~\eqref{bosonizationDC} and below] from that of the transmission and reflection coefficient in the fermionic representation. The possible non-zero commutators among $a_{k}$ and $b_{k}$ operators are $[a_{k_1}, a^{\dagger}_{k_2}]= 2\pi  \delta(k_1-k_2)$ and $[b_{k_1}, b^{\dagger}_{k_2}]= 2\pi  \delta(k_1-k_2)$. The canonical commutation relation requires 
\begin{align}\label{commutation}
1-|R_{\mathrm{L}}(k)|^2= &|S_{\mathrm{R}}(k)|^2 , \nonumber \\
1-|R_{\mathrm{R}}(k)|^2= &|S_{\mathrm{L}}(k)|^2  . 
\end{align}
Due to the continuity of the charge current $\sim \partial_t \phi$ , for $k \rightarrow 0$ we obtain
\begin{align}~\label{ccurrentcontinuation}
\sqrt{K_{\mathrm{L}}}[1+R_{\mathrm{L}} (k)]= & \sqrt{K_{\mathrm{R}}} S_{\mathrm{L}}(k) , \nonumber \\ 
\sqrt{K_{\mathrm{R}}}[1+R_{\mathrm{R}} (k)]= & \sqrt{K_{\mathrm{L}}} S_{\mathrm{R}}(k) . 
\end{align}
Similarly due to the continuity of the energy current $\sim v \partial_t \phi \partial_x \phi$, it follows:
\begin{align}~\label{ecurrentcontinuation}
1-|R_{\mathrm{L}}(k)|^2= &|S_{\mathrm{L}}(k)|^2 , \nonumber \\
1-|R_{\mathrm{R}}(k)|^2= &|S_{\mathrm{R}}(k)|^2 . 
\end{align}
One of the real solutions of Eqs.~\eqref{commutation},~\eqref{ccurrentcontinuation}, and~\eqref{ecurrentcontinuation}, which turns out to correspond to the maximal conductance, is
\begin{align}\label{maxtransmissionRS}
R_{\mathrm{L}}(k=0^{+})=-R_{\mathrm{R}}(k=0^{+})=&\frac{K_{\mathrm{R}}-K_{\mathrm{L}}}{K_{\mathrm{L}}+K_{\mathrm{R}}} , \nonumber \\
S_{\mathrm{R}}(k=0^{+})=S_{\mathrm{L}}(k=0^{+})=&\frac{2\sqrt{K_{\mathrm{L}}K_{\mathrm{R}}}}{K_{\mathrm{L}}+K_{\mathrm{R}}} .
\end{align}
Another real solution reads $R_{\mathrm{L}}(k=0^{+})=R_{\mathrm{R}}(k=0^{+})=-1$, and $S_{\mathrm{L}}(k=0^{+})=S_{\mathrm{R}}(k=0^{+})=0$, which corresponds to  zero conductance.

Let us calculate the charge fluctuation in the infinite configuration. We first compute the density-density correlation function in the left region for large $|x_1-x_2|$, 
\begin{align}
&\left\langle  n(x_1) n(x_2) \right\rangle \nonumber - \left\langle n_{x_1} \right \rangle \left \langle n_{x_2} \right\rangle \\
=&\left\langle \partial_{x_1} \phi(x_1) \partial_{x_2} \phi(x_2) \right\rangle/\pi^2 \nonumber+ \ldots \\
=&\frac{K_{\mathrm{L}}}{2\pi^2(x_1-x_2)^2}+\frac{K_{\mathrm{L}}(R_{\mathrm{L}}+R^{*}_{\mathrm{L}})}{4\pi^2(x_1+x_2)^2}+ \ldots,
\end{align}
where the oscillatory and higher order terms have been neglected. Summing over $x_1, x_2$ in a segment, the charge fluctuation is obtained as in Eq.~(\ref{variancefromcorrelation}). The first term alone contributes to the leading behavior of the charge fluctuation of a segment of length $l$ in the same way as a homogeneous chain but with a prefactor of $K_{\mathrm{L}}$; so its leading behavior is $\frac{K_{\mathrm{L}}}{\pi^2}\ln(l)$.  This can be seen by assuming Eqs.~\eqref{maxtransmissionRS} and setting $K_{\mathrm{R}}=K_{\mathrm{L}}$. The summation or integration (with a UV cut-off) of the second term contributes $\frac{K_{\mathrm{L}}(R_{\mathrm{L}}+R^{*}_{\mathrm{L}})}{4\pi^2}\ln(l)$. Therefore, in total 
\begin{align}\label{singlechannelchargevariance}
\left\langle N_{\mathrm{seg}}^2 \right\rangle-\left\langle N_{\mathrm{seg}} \right\rangle^2 = \frac{G_{\mathrm{h}}+ G}{2\pi^2}\ln(l)+ \ldots, 
\end{align}
where $G_{\mathrm{h}}=K_{\mathrm{L}}$ and $G$ is given by Eq.~\eqref{bosonizationDC}.

\section{} \label{app_C}

In this appendix, we present additional data for the QPC model Eq.~\eqref{HQPC} and the interacting-lead model with abruptly changing interaction Eqs.~\eqref{HIL} and \eqref{abruptV}.

\begin{figure*}[tbp]
\includegraphics[width=\textwidth]{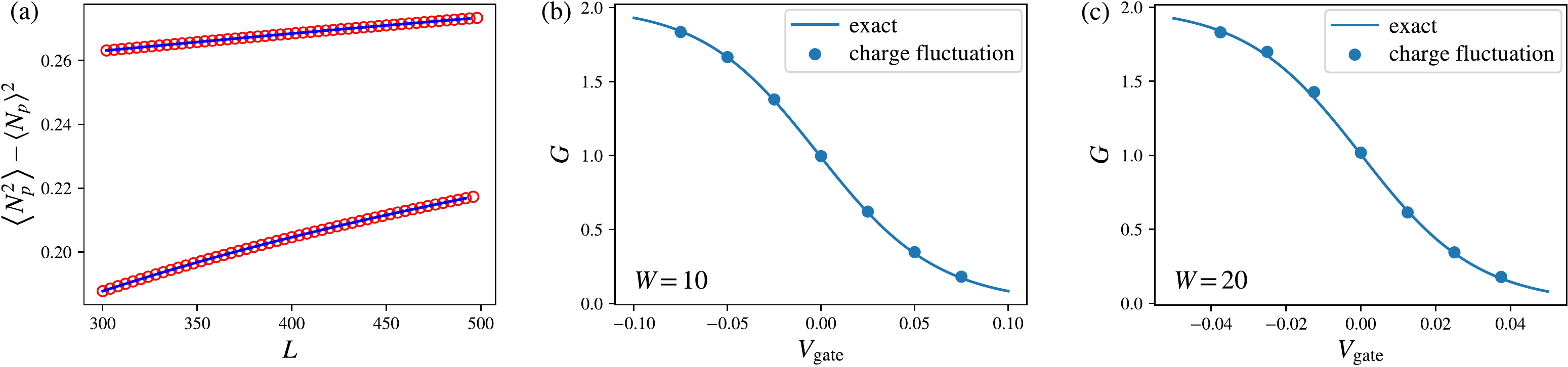}
\caption{Dc conductance of the noninteracting QPC model [Eq.~\eqref{HQPC} with $U=0$]. The model parameters are the same as in Fig.~\ref{fig:QPC}, except that different half widths $W$ are investigated. (a) Raw data of charge fluctuations for one spin component with $W=20, V_{\mathrm{gate}}=0$. The open red circles are data obtained from exact diagonalization, the blue lines are fitting results using all data and Eq.~\eqref{Gfitansatz}. (b), (c) $G$ fitted from charge fluctuations following the example of (a) and averaging the results from two branches. For larger QPC size ($W$), we expect using larger total size is required for keeping the accuracy. With the same total sizes, the accuracy is lower for larger $W$. For $W=20$, we see small but clear deviation from the exact values.}\label{fig:QPCaddition}
\end{figure*}

In Fig.~\ref{fig:QPCaddition}, we show size-dependent data for the charge fluctuation, a fitting to these and the extracted $G$ of larger systems. We consider the noninteracting case. This allows us to use exact diagonalization to obtain charge fluctuation data, which is computationally less demanding than DMRG. Hence, we can obtain more data points and use larger systems to illustrate that the fitting works well; see Fig.~\ref{fig:QPCaddition}(a). However, using more data points and larger systems provides no significant further improvement for the fitting  accuracy compared to the data shown in the main text. The accuracy is instead controlled by the ratio of device size over the total system size. In Figs.~\ref{fig:QPCaddition}(b) and ~\ref{fig:QPCaddition}(c), we show that for fixed total system sizes, larger QPC device sizes lead to an increasing error. We do so by comparing to the exact result at vanishing interaction. The results obtained for the larger half width ($W=20$) QPC have errors comparable to the symbol size.

\begin{figure}[tbp]
\includegraphics[width=0.9\columnwidth]{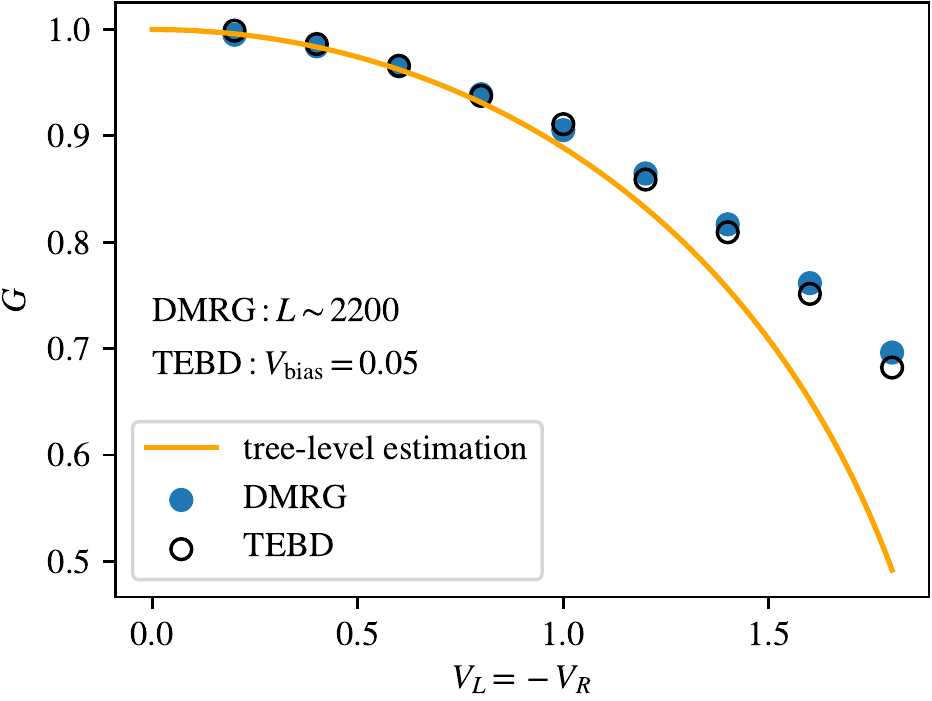}
\caption{The same setup as case studied in  Fig.~\ref{fig:xxzabrupt}(a) with $N_l=0$ and considering different $V_{\mathrm{L}}$. The tree-level estimation is given by $\lambda=(v_{\mathrm{L}}-v_{\mathrm{R}})/(2\pi)$ with the Bethe ansatz formula $v_{\mathrm{L},\mathrm{R}}=\pi\sqrt{1-V^2_{\mathrm{L},\mathrm{R}}/4}/\arccos (V_{\mathrm{L},\mathrm{R}}/2)$ and $G=1-[2\pi\lambda/(v_{\mathrm{L}}+v_{\mathrm{R}})]^2$.}\label{fig:xxzabruptaddition}
\end{figure}

In Fig.~\ref{fig:xxzabruptaddition}, we show more data for the case  of the interacting-lead model with abruptly changing interaction [Eqs.~\eqref{HIL} and \eqref{abruptV}]. 
As discussed in the main text,  the tree-level theory predicts no transport resonance and the bare backscattering does not renormalize. The latter fact motivates a quantitative comparison between the backscattering estimated by the tree-level theory~\cite{conductingfixedpoint1, conductingfixedpoint2} and our data for $G$. We find that the results agree well in the weak coupling limit (small $V_{\mathrm{L}}$ and $V_{\mathrm{R}}$), where the tree-level theory is expected to work quantitatively. We note that this kind of comparison goes beyond the one presented in the related literature~\cite{conductingfixedpoint1, conductingfixedpoint2}.

\bibliography{ref}

\end{document}